\documentclass[12pt,letterpaper,psamsfonts,reqno]{amsart}

\usepackage[scale=0.85,centering,foot=1cm]{geometry}
\usepackage{graphicx}
\usepackage{amsmath,amssymb,amsthm,bm}
\usepackage{epsfig,color,cases,cite,url}

\newcommand{\Integer}{\mathbb{Z}}
\newcommand{\Natural}{\mathbb{Z}_{\geq 0}}
\newcommand{\Naturalstar}{\mathbb{Z}_{> 0}}
\newcommand{\Real}{\mathbb{R}}

\newcommand{\norm}[1]{\|{#1}\|}

\DeclareMathOperator{\argmin}{arg\,min}

\DeclareMathOperator{\relinterior}{ri}
\DeclareMathOperator{\interior}{int}

\DeclareMathOperator{\sign}{sgn}

\DeclareMathOperator{\lev}{lev_{\leq 0}}

\DeclareMathOperator{\abs}{abs}
\DeclareMathOperator{\supp}{supp}

\theoremstyle{definition}
\newtheorem{theorem}{Theorem}

\newtheorem{fact}{Fact}

\newtheorem{lemma}{Lemma}

\newtheorem{remark}{Remark}
\newtheorem{definition}{Definition}

\begin{document}

\title{Online Sparse System Identification and Signal Reconstruction
  using Projections onto Weighted $\ell_1$ Balls}

\author{Yannis Kopsinis}
\author{Konstantinos Slavakis}
\author{Sergios Theodoridis}

\thanks{Y.~Kopsinis and S.~Theodoridis are with the
University of Athens, Department of Informatics and Telecommunications,
Ilissia, Athens 15784, Greece. Emails: \url{kopsinis@ieee.org},
\url{stheodor@di.uoa.gr}. Tel: +30.210.727.5328, Fax:
+30.210.727.5337.}
\thanks{K.~Slavakis is with the University of Peloponnese, Department of
Telecommunications Science and Technology, Karaiskaki St., Tripolis
22100, Greece. Email: \url{slavakis@uop.gr}. Tel: +30.2710.37.2204,
Fax: +30.2710.37.2160.}


\maketitle

\begin{abstract}
This paper presents a novel projection-based adaptive algorithm for
sparse signal and system identification. The sequentially observed
data are used to generate an equivalent sequence of closed convex
sets, namely hyperslabs. Each hyperslab is the geometric equivalent of
a cost criterion, that quantifies ``data mismatch''. Sparsity is
imposed by the introduction of appropriately designed weighted
$\ell_1$ balls. The algorithm develops around projections onto the
sequence of the generated hyperslabs as well as the weighted $\ell_1$
balls. The resulting scheme exhibits linear dependence, with respect to the
unknown system's order, on the number of multiplications/additions and
an $\mathcal{O}(L\log_2L)$ dependence on sorting operations, where $L$
is the length of the system/signal to be estimated. Numerical results
are also given to validate the performance of the proposed method
against the LASSO algorithm and two very recently developed adaptive
sparse LMS and LS-type of adaptive algorithms, which are considered to
belong to the same algorithmic family.
\end{abstract}

\section{Introduction}\label{sec:intro}

Sparsity is the key characteristic of systems whose impulse response
consists of only a few nonzero coefficients, while the majority of
them retain values of negligible size.  Similarly, any signal
comprising a small number of nonzero samples is also characterized as
being a sparse one. The exploitation of sparsity has been attracting
recently an interest of exponential growth under the Compressed
Sensing (CS) framework \cite{CandesRombergTao06, Donoho2006,
  CandesICM06}. In principle, CS allows the estimation of sparse
signals and systems using fewer measurements than those previously
thought to be necessary. More importantly,
identification/reconstruction is realized with efficient constrained
minimization schemes. Indeed, it has been shown that sparsity is
favored by $\ell_1$ constrained solutions
\cite{CandesTao05,Baraniuk07}.

With only a few recent exceptions, i.e., \cite{ChenHero09,
  Angelosante09, Kalouptsidis.Sparse.09, myyy.soft.thres.icassp10,
  Slav.Kops.Theo.ICASSP10}, the majority of the proposed, so far, CS
techniques are appropriate for batch mode operation. In other words,
one has to wait until a fixed and predefined number of measurements is
available prior to application of CS processing methods, in order to
recover the corresponding signal/system estimate. Dynamic online
operation for updating and improving estimates, as new measurements
become available is not feasible by batch processing methods. The
development of efficient, online adaptive CS techniques is of great
importance, especially for cases where the signal or system under
consideration is time-varying and/or if the available storage
resources are limited.

The basic idea in \cite{ChenHero09, Angelosante09} is to use
\textit{$\ell_1$ regularization}, i.e., to add to a standard linear or
quadratic loss function an extra penalty term expressed by means of
the well-known $\ell_1$ norm of the unknown system/signal coefficients. Such
an approach has been adopted for the classical LMS \cite{ChenHero09},
and for  the LS type \cite{Angelosante09} minimization problems. The
resulting recursions for the time update use the current estimate and
the information residing in the subgradient of the cost function (due
to the non-differentiability of the $\ell_1$ norm) to provide the next
estimate.

This paper evolves along a different rationale compared to
\cite{ChenHero09, Angelosante09}, and introduces a projection-based
algorithm for sparse system identification and sparse signal
reconstruction. The kick-off point is the set theoretic estimation
approach, e.g., \cite{CombettesFoundations}. Instead of a single
optimum, we search for a set of points that are in {\it agreement}
with the available information, which resides in the training data set
(measurements) as well as in the available constraints (the $\ell_1$
ball, in this case). To this end, as each new set of measurements is
received, a closed convex set is constructed, which defines the region
in the solution space that is in ``agreement'' with the current
measurement.  In context of the current paper, the shape of these
convex sets is chosen to be a hyperslab.  The resulting problem is a
convex feasibility task, with an \textit{infinite} number of convex
constraints. The fundamental tool of projections onto closed convex
sets is used to tackle the problem, following the recent advances on
adaptive projection algorithms \cite{YamadaOguraAPSMNFAO,
  KostasAPSMatNFAO, KostasClassifyTSP}. Instead of using the
information associated with the subgradient of the $\ell_1$ norm, the
$\ell_1$ constraint is imposed on our solution via the exact
projection mapping onto a weighted $\ell_1$ ball. The algorithm
consists of a sequence of projections onto the generated hyperslabs as
well as the weighted $\ell_1$ balls.  The associated complexity is of
order $\mathcal{O}(qL)$ multiplications/additions and
$\mathcal{O}(L\log_2L)$ sorting operations, where $L$ is the length of
the system/signal to be identified and $q$ is a user-defined
parameter, that controls convergence speed and it defines the number
of measurements that are processed, concurrently, at each time
instant.  The resulting algorithm enjoys a clear geometric
interpretation.

The paper is organized as follows. In Section~\ref{sec:model} the
problem under consideration is described and in
Section~\ref{sec:online.estimation} some definitions and related
background are provided. Section~\ref{sec:algo} presents the proposed
algorithm. The derivation and discussion of the projection mapping
onto the weighted $\ell_1$ ball are treated in
Section~\ref{sec:projections}. The adopted mechanism for weighting the
$\ell_1$ ball is discussed in Section~\ref{sec:weights.specs}. In
Section~\ref{sec:convergence}, the convergence properties of the
algorithm are derived and discussed. It must be pointed out that this
section comprises one of the main contributions of the paper, since
the existing, so far, theory cannot cover the problem at hand and has
to be extended.  In Section~\ref{sec:simulation.examples}, the
performance of the proposed algorithmic scheme is evaluated for both,
time-invariant and time-varying
scenarios. Section~\ref{sec:sensitivity} addresses issues related to
the sensitivity of the methods, used in the simulations, to non-ideal
parametrization and, finally, the conclusions are provided in
Section~\ref{sec:conclusions}. The Appendices offer a more detailed
tour to the necessary, for the associated theory, proofs.

\section{Problem description}\label{sec:model}

We will denote the set of all integers, non-negative integers, positive
integers, and real numbers by $\Integer,\ \Natural$, $\Naturalstar$,
and $\Real$, respectively. Given two integers $j_1, j_2\in\Integer$,
such that $j_1\leq j_2$, let $\overline{j_1, j_2}:= \{j_1, j_1+1,
\ldots,j_2\}$.

The stage of discussion will be the Euclidean space $\Real^L$, of
dimension $L\in\Naturalstar$. Its norm will be denoted by
$\norm{\cdot}$. The superscript symbol $(\cdot)^T$ will stand for vector
transposition. The $\ell_1$ norm of a vector $\bm{h}= [h_1,\ldots,
h_L]^T \in\Real^L$ is defined as the quantity $\norm{\bm{h}}_{\ell_1} :=
\sum_{i=1}^L |h_i|$. The support of a vector $\bm{h}$ is defined as
$\supp(\bm{h}):= \{i\in\overline{1,L}:\ h_i \neq 0\}$. The
$\ell_0$-norm of $\bm{h}$ is defined as the cardinality of its
support, i.e., $\norm{\bm{h}}_{\ell_0}:= \# \supp(\bm{h})$.

Put in general terms, the problem to solve is to estimate a vector
$\bm{h}_*$, based on measurements that are sequentially generated by
the (unknown) linear regression model:
\begin{equation}
y_n = \bm{x}_n^T\bm{h}_*  + v_n,\forall n\in\Natural,
\label{eq:regression_Model}
\end{equation}
where the model outputs $(y_n)_{n\in\Natural} \subset\Real$ and the
model input vectors $(\bm{x}_n)_{n\in\Natural} \subset \Real^L$
comprise the measurements and $(v_n)_{n\in\Natural}$ is the noise
process. Furthermore, the unknown vector $\bm{h}_*$ is $S$-sparse,
meaning that it has $S$ non-zero terms only, with $S$ being small
compared to $L$, i.e., $S:= \norm{\bm{h}_*}_{\ell_0} \ll L$.

For a finite number of measurements $N$, the
previous data generation model can be written compactly in the
following matrix-vector form,
\begin{equation}
\bm{y}=\bm{X}\bm{h}_*+\bm{v},
\label{eq:regression_Modelmatrixvector}
\end{equation}
where the input matrix $\bm{X} \in \Real^{N\times L}$ has as its rows
the input measurement vectors, $\bm{y} := [y_1,y_2,\ldots, y_N]^T$,
and $\bm{v} := [v_1,v_2,\ldots, v_N]^T$.

Depending on the physical quantity that $\bm{h}_*$ represents, the
model in \eqref{eq:regression_Modelmatrixvector} suits to both sparse
signal reconstruction and linear sparse system identification:

\begin{enumerate}
\item
\textbf{Sparse signal reconstruction problem:} The aim is to estimate
an unknown sparse signal, $\bm{h}_*$, based on a set of measurements
(training data), that are obtained as inner products of the unknown
signal with appropriately selected input vectors, $\bm{x}_n$,
according to \eqref{eq:regression_Model}. The elements of the input
vectors are often selected to be independent and identically
distributed (i.i.d.) random variables following, usually, a zero-mean
normal or a Bernoulli distribution \cite{Baraniuk07}.

\item \textbf{System identification problem:} The unknown sparse
  system with impulse response $\bm{h}_*$ is probed with an input
  signal $x_n$, $n\in\Natural$ yielding the output values $y_n$ as the
  result of convolution of the input signal with the (unknown) impulse
  response of the system. In agreement to the model of
  \eqref{eq:regression_Model}, the measurement (input) vector, at time
  $n$, is given by $\bm{x}_n := [x_n,x_{n-1},\ldots, x_{n-L+1}]^T$. In
  the matrix-vector formulation, and for a finite number of
  measurements, the corresponding measurement matrix $\bm{X}$ is a
  (partial) Toeplitz one having as entries the elements
  $\bm{T}_{i,j}=x_{i+L-j}$, where $i\in\overline{1,N}$ and
  $j\in\overline{1,L}$. The input signal vector, $\bm{x}$, usually
  consists of i.i.d.\ normally distributed samples. The study of
  Toeplitz matrices, with respect to their potential to serve as CS
  measurement matrices, has been recently intensified, e.g.,
  \cite{Holger:2009,Bajwa:SSP2007}, partially due to their importance
  in sparse channel estimation applications \cite{Bajwa:2010}.
\end{enumerate}

A batch approach to estimating a sparse $\bm{h}_*$ based on a limited
number of measurements $N<L$, is provided by the Least-Absolute
Shrinkage and Selection Operator (LASSO):
\begin{equation}
    \bm{h}_*=\argmin_{\bm{h}:\ \|\bm{h}\|_{\ell_1} \leq \delta}
    \|\bm{X} \bm{h}-\bm{y}\|^2.
\end{equation}
In this case, 
$\bm{h}_*$ is assumed to be stationary and the total number of
measurements, $N$, needs to be available prior to solution of the
LASSO task.

In the current study, we will assume that $\bm{h}_*$ is not only
sparse but it is also allowed to be time-varying.
This poses certain distinct differences with regard
to the standard compressive sampling scenario. The major objective is
no longer the estimate of the sparse signal or system, based on a
limited number of measurements. The additional requirement,
which is often more hard to cope with, is the capability of the
estimator to track possible variations of the unknown signal or
system. Moreover, this has to take place at an affordable
computational complexity, as required by most real time applications,
where online adaptive estimation is of interest. Consequently, the
batch \textit{sparsity aware} techniques developed under the CS
framework, solving LASSO or one of its variants, become unsuitable under
time-varying scenarios.  The focus now becomes to develop techniques
that a) exploit the sparsity b) exhibit fast convergence to error
floors that are as close as possible to those obtained by their batch
counterparts c) offer good tracking performance and d) have low
computational demands in order to meet the stringent time constraints
that are imposed by most real time operation scenarios.

\section{Online estimation under the sparsity
  constraint}\label{sec:online.estimation}

The objective of online techniques is the generation of a sequence of
estimates, $(\bm{h}_n)_{n\in\Natural}$, as time, $n$, evolves, which
converge to a value that \textit{``best approximates''}\/, in some sense,
the unknown sparse vector $\bm{h}_*$. The classical approach to this
end is to adopt a loss function and then try to minimize it in a time
recursive manner. A more recent approach is to achieve the goal via
set theoretic arguments by exploiting the powerful tool of projections.

\subsection{Loss function minimization approach.}

A well-known approach to quantify the ``best approximation'' term is
the minimization of a user-defined loss function
\begin{equation}
\forall n\in\Natural, \forall \bm{h}\in\Real^L, \quad
\Theta_n(\bm{h}):= \mathcal{L}_r^{(n)}(\bm{h}) + \gamma_n
\mathcal{L}_s^{(n)}(\bm{h}), \label{cost.function}
\end{equation}
where $\mathcal{L}_r^{(n)}$ is computed over the training (observed)
data set and accounts for the data mismatch, between measured and
desired responses,
and $\mathcal{L}_s^{(n)}$ accounts for the ``size'' of the solution, and
in the current context is the term that imposes \textit{sparsity}. The
sequence of user-defined parameters $(\gamma_n)_{n\in\Natural}$
accounts for the relative contribution of $\mathcal{L}_r^{(n)},
\mathcal{L}_s^{(n)}$ to the cost in \eqref{cost.function}. Usually,
both functions $\mathcal{L}_r^{(n)},\mathcal{L}_s^{(n)}$ are chosen to
be convex, due to the powerful tools offered by the convex analysis
theory.

For example, the study in \cite{ChenHero09} chooses
$\mathcal{L}_r^{(n)}(\bm{h}) := \frac{1}{2} |y_n - \bm{h}^T
\bm{x}_n|^2$, where
$\mathcal{L}_s^{(n)}(\bm{h}):= \norm{\bm{h}}_{\ell_1}$,
$\bm{h}\in\Real^L$, in order to obtain the ZA-LMS algorithm. The
RZA-LMS scheme is obtained in \cite{ChenHero09} when setting
$\mathcal{L}_s^{(n)}(\bm{h}):= \sum_{i=1}^L
\log(1+\frac{|h_i|}{\eta})$, $\bm{h}\in\Real^L$, while keeping the
same $\mathcal{L}_r^{(n)}$. In \cite{Angelosante09}, the sum Least
Squares with a forgetting factor $\beta$ is used in place of
$\mathcal{L}_r^{(n)}$ and the $\ell_1$ norm in $\mathcal{L}_s^{(n)}$.

\subsection{Set theoretic approach.}

In this paper, a different path is followed. Instead of attempting to
minimize, recursively, a cost function that is defined over the entire
observations' set, our goal becomes to find a {\it set} of solutions
that is in \textit{agreement} with the available observations as well as
the constraints. To this end, at each time instant, $n$, we require
our estimate $\bm{h}_n$ to lie
within an appropriately defined closed convex set, which is a subset
of our solutions space and it is also known as {\it property set}. Any
point that lies within this set is said to be in agreement with the
current measurement pair $(\bm{x}_n,y_n)$. The ``shape'' of the
property set is dictated by a ``local'' loss function, which is
assumed to be convex. In the context of the current paper, we have
adopted property sets that are defined by the following criterion
\begin{equation}
S_n[\epsilon] := \{\bm{h}\in\Real^L:\ |\bm{h}^T \bm{x}_n - y_n| \leq
\epsilon \},\ n\in\Natural,\label{Hyperslab}
\end{equation}
for some user-defined tolerance $\epsilon\geq 0$. Such criteria have
extensively been used in the context of robust statistics cost
functions. Eq.~\eqref{Hyperslab} defines a \textit{hyperslab}, which is
indeed a closed convex set. Any point that
lies in the hyperslab generated at time $n$ is in agreement with the
corresponding measurement at the specific time instance. The
parameter $\epsilon$ determines the width of the
hyperslabs. Fig.~\ref{fig:projections} shows two hyperslabs defined at
two successive instants, namely, $n$ and $n-1$.

Having associated each measurement pair with a hyperslab, our goal,
now, becomes to find a point in $\Real^L$ that lies in the
intersection of these hyperslabs, provided that this is nonempty. We
will come back to this point when discussing the convergence issues of
our algorithm. For a recent review of this algorithmic family the
reader may consult \cite{TheodoridisSlavYamReview}.

To exploit sparsity, we adopt the notion of the weighted $\ell_1$
norm. Given a vector $\bm{w}_n\in\Real^L$ with positive components,
i.e., $w_{n,i} > 0$, $\forall i\in\overline{1,L}$, the \textit{weighted
  $\ell_1$ ball of radius $\delta>0$} is defined as \cite{CandesWakinBoyd08}
\begin{equation}
B_{\ell_1}[\bm{w}_n,\delta] := \{\bm{h}\in\Real^L:\ \sum_{i=1}^{L}
 w_{n,i} |h_i| \leq \delta\}.\label{l1.ball}
\end{equation}
For more flexibility, we let the weight vector depend on the time
instant $n$, hence the notation $\bm{w}_n$ has been adopted. We will
see later on that such a strategy speeds up convergence and decreases
the misadjustment error of the algorithm. The well-known \textit{unweighted
$\ell_1$ ball} is nothing but $B_{\ell_1}[\bm{1},\delta]$, where
$\bm{1}\in\Real^L$ is a vector with $1$s in all of its
components. Note that all the points that lie inside a weighted
$\ell_1$ norm form a closed convex set.

Having defined the weighted $\ell_1$ ball, which is the sparsity related
constraint, our task now is to search for a point $\bm{h}$ in
$\Real^L$ that lies in the intersection of the hyperslabs as well as
the weighted $\ell_1$ balls, i.e., for some $z_0\in\Natural$,
\begin{equation}
\text{find an}\ \bm{h} \in \bigcap_{n\geq z_0} \left(S_n[\epsilon] \cap
 B_{\ell_1}[\bm{w}_n,\delta]\right).\label{the.problem}
\end{equation}
As it will become clear later on, when discussing the convergence
issues of the algorithm, the existence of $z_0$ in (\ref{the.problem})
allows for a finite number of property sets not to share intersection
with the rest.

\section{Proposed algorithmic framework}\label{sec:algo}

The solution to the problem of finding a point lying in the
intersection of a number of closed convex sets has been developed in
the context of the classical POCS theory \cite{BregmanPocs,
  PolyakPocs, BauschkeBorwein, StarkYangBook}, in the case where there
is a finite number of sets, and its recent extension, that deals with
an \textit{infinite number} of sets, originally proposed in
\cite{YamadaOguraAPSMNFAO}. The basic idea is very elegant: Keep
projecting, according to an appropriate rule, on the involved convex
sets; then this sequence of projections will, finally, take you to a
point in their intersection. Hence, for our problem, metric projection
mapping operators for, both, the hyperslabs as well as the weighted
$\ell_1$ balls have to be used. Projection operators for hyperslabs
are already known and widely used, e.g., \cite{ksi.Beam.TSP,
  TheodoridisSlavYamReview}. The metric projection mapping onto a
weighted $\ell_1$ norm will be derived here, and it was presented for
a first time, to the best of our knowledge, in
\cite{Slav.Kops.Theo.ICASSP10}.

\begin{figure}[t]
\begin{minipage}[b]{1.0\linewidth}
  \centering
 \centerline{\epsfig{figure= 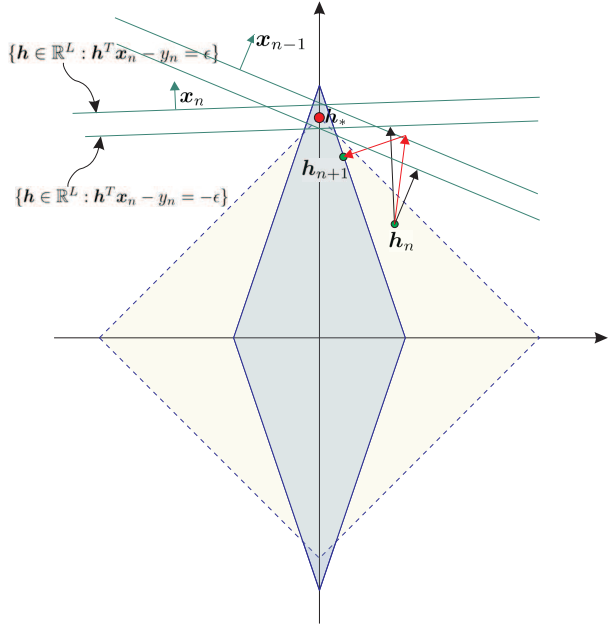,width=8cm}}
\end{minipage}
\caption{The $\ell_1$ ball is shown with dotted lines. At time $n$,
  the estimate $\bm{h}_{n}$ is available. For $q=2$, two hyperslabs are
  involved in the update recursion, those associated with time
  instants $n$ and $n-1$. The new update, $\bm{h}_{n+1}$, results by
  first projecting $\bm{h}_{n}$ onto the hyperslabs, then combining the
  resulting projections and finally projecting onto the weighted $\ell_1$
  norm, that is defined at time $n$ and which is drawn using the full
  line.}\label{fig:projections}
\end{figure}

Each time instant, $n$, a new pair of training data $(\bm{x}_n,y_n)$
becomes available, and a corresponding hyperslab is formed according
to \eqref{Hyperslab}. This is used to update the currently available
estimate $\bm{h}_n$.  However, in order to speed up convergence, the
update mechanism can also involve previously defined hyperslabs; for
example, the hyperslabs formed at time instants $\overline{n-q+1,n}$,
for some $q\in\Naturalstar$. Then, in order to obtain $\bm{h}_{n+1}$,
an iteration scheme consisting of three basic steps, is adopted: a)
the current estimate $\bm{h}_n$ is projected onto each one of the $q$
hyperslabs, b) these projections are in turn combined as a weighted
sum and c) the result of the previous step is subsequently projected
onto the weighted $\ell_1$ ball. This is according to the concepts
introduced in \cite{YamadaOguraAPSMNFAO} and followed in
\cite{KostasAPSMatNFAO, KostasClassifyTSP, ksi.Beam.TSP}. Schematically,
the previous procedure is illustrated in Fig.~\ref{fig:projections},
for the case of $q=2$.

In detail, the algorithm is mathematically described as follows:

\noindent{\bf Algorithm.} Let $q\in\Naturalstar$, and define the following
sliding window on the time axis, of size at most $q$ (to account for
the initial period where $n<q-1$), in order to indicate the hyperslabs
to be considered at each time instant:
\begin{equation*}
\quad \mathcal{J}_n := \overline{\max\{0,n-q+1\},n}, \quad \forall
n\in\Natural.
\end{equation*}
For each $n$, define the set of weights
$\{\omega_j^{(n)}\}_{j\in\mathcal{J}_n} \subset(0,1]$ such that
  $\sum_{j\in\mathcal{J}_n} \omega_j^{(n)}=1$. Each $\omega_j^{(n)}$
  quantifies the contribution of the $j$-th hyperslab into the
  weighted combination of all the hyperslabs that are represented
  indicated in $\mathcal{J}_n$.

Given an arbitrary initial point $\bm{h}_0\in\Real^L$, the following
recursion generates the sequence of estimates
$(\bm{h}_n)_{n\in\Natural}$; $\forall n\in\Natural$,
\begin{equation}
\bm{h}_{n+1} :=
P_{B_{\ell_1}[\bm{w}_n,\delta]}\left(
\bm{h}_n + \mu_n \left(\sum_{j\in\mathcal{J}_n} \omega_j^{(n)}
P_{S_j[\epsilon]}(\bm{h}_n) - \bm{h}_n \right)\right),\label{algo}
\end{equation}
where $P_{S_j[\epsilon]}$ and $P_{B_{\ell_1}[\bm{w}_n,\delta]}$ denote
the metric projection mappings onto the hyperslab, defined by the
$j$-th data pair, and onto the, (currently available) weighted
$\ell_1$ ball, respectively. As it will be shown in the analysis of
the algorithm in Appendix~\ref{app:analysis.algo}, in order to
guarantee convergence, the extrapolation parameter $\mu_n$ takes
values within the interval $(0,2\mathcal{M}_n)$, where $\mathcal{M}_n$
is computed by
\begin{equation}
\mathcal{M}_n:=%
\begin{cases}
\frac{ \sum_{j\in\mathcal{J}_n} \omega_j^{(n)}
 \norm{P_{S_j[\epsilon]}(\bm{h}_n) - \bm{h}_n}^2} {
 \norm{\sum_{j\in\mathcal{J}_n} \omega_j^{(n)}
   P_{S_j[\epsilon]}(\bm{h}_n) - \bm{h}_n}^2}, &
\text{if}\ \sum_{j\in\mathcal{J}_n} \omega_j^{(n)}
   P_{S_j[\epsilon]}(\bm{h}_n) \neq \bm{h}_n,\\
1, & \text{otherwise}. \label{Mn}
\end{cases}
\end{equation}
Notice that the convexity of the function $\norm{\cdot}^2$ implies
that $\mathcal{M}_n \geq 1$, $\forall n\in\Natural$.

It is interesting to point out that the algorithm is compactly encoded
into a single equation! Also, note that projection onto the $q$
hyperslabs can take place concurrently and this can be exploited if
computations are carried our in a parallel processing
environment. Moreover, $q$ can be left to vary from iteration to
iteration. The dependence of the performance of the algorithm on $q$
will be discussed in Section~\ref{sec:simulation.examples}.

It turns out that the projection mappings involved in \eqref{algo} and
\eqref{Mn} have computationally simple forms and are given in
\eqref{ProjHyperslab} and Section~\ref{sec:proj.onto.ball}. The
algorithm amounts to a computational load of order $\mathcal{O}(qL)$
multiplications/additions and $\mathcal{O}(L\log_2L)$ sorting
operations. The dependence on $q$ is relaxed in a parallel processing
environment.

Having disclosed the algorithmic scheme for the update of our estimate
at each iteration step, as measurements are received sequentially,
there are a number of issues, yet, to be resolved. First, the involved
projection mappings have to be explicitly provided/derived.  Second, a
strategy for the selection of the weights in the weighted $\ell_1$
norm need to be decided. Third, the convergence of the algorithm has
to be established. Although the algorithm stands on the shoulders of
the theory developed in previously published papers, e.g.,
\cite{YKostasY, YamadaOguraAPSMNFAO, KostasAPSMatNFAO}, the developed,
so far, theory is not enough to cover the current algorithm. Since we
do not use the $\ell_1$ norm, but its weighted version, the projection
mapping $P_{B_{\ell_1}[\bm{w}_n,\delta]}$ in \eqref{algo} is
\textit{time-varying} and it also depends on the obtained
estimates. Convergence has to be proved for such a scenario and this
is established in Appendix~\ref{app:analysis.algo}.

\section{Projections onto Closed Convex Sets}\label{sec:projections}

A subset $C$ of $\Real^L$ will be called convex if every line segment
$\{\lambda \bm{h} + (1-\lambda)\bm{h}': \lambda\in [0,1] \}$, with
endpoints any $\bm{h}, \bm{h}' \in C$, lies in $C$.

Given any set $C\subset\Real^L$, define the \textit{(metric) distance
  function $d(\cdot,C): \Real^L\rightarrow\Real$ to $C$} as follows:
$\forall \bm{x}\in\Real^L$, $d(\bm{x},C):= \inf\{\norm{\bm{x}-
  \bm{f}}:\ \bm{f}\in C\}$. If we assume now that $C$ is closed and
convex, then the \textit{(metric) projection onto $C$} is defined as
the mapping $P_C: \Real^L \rightarrow C$ which maps to an
$\bm{x}\in\Real^L$ the unique $P_C(\bm{x})\in C$ such that
$\norm{\bm{x}-P_C(\bm{x})} = d(\bm{x},C)$.

\subsection{Projecting onto a hyperslab.}

The metric projection operator $P_{S_n[\epsilon]}$ onto the
hyperslab \eqref{Hyperslab} takes the following simple analytic form
\cite{BauschkeBorwein, StarkYangBook}:
\begin{equation}
\forall \bm{h}\in\Real^L, \quad P_{S_n[\epsilon]}(\bm{h}) = \bm{h}
+ \begin{cases} \frac{y_n - \epsilon - \bm{h}^T \bm{x}_n}
  {\norm{\bm{x}_n}^2}\bm{x}_n, & \text{if}\ y_n - \epsilon > \bm{h}^T
  \bm{x}_n,\\ \bm{0}, & \text{if}\ |\bm{h}^T \bm{x}_n - y_n |\leq
  \epsilon,\\ \frac{y_n + \epsilon - \bm{h}^T \bm{x}_n}
          {\norm{\bm{x}_n}^2}\bm{x}_n, & \text{if}\ y_n + \epsilon <
          \bm{h}^T \bm{x}_n.
\end{cases}\label{ProjHyperslab}
\end{equation}

\subsection{Projecting onto the weighted $\ell_1$
 ball.}\label{sec:proj.onto.ball}

\begin{figure}[t]
\center
\scalebox{0.5}{\input{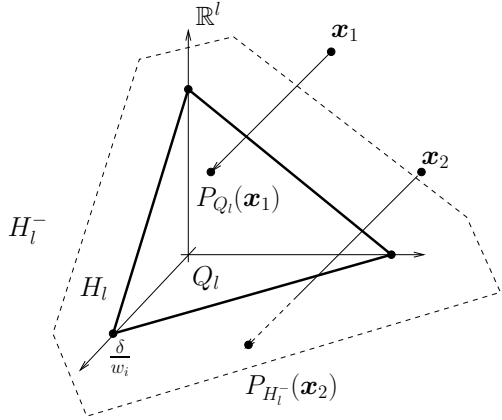}} \caption{This
  figure illustrates the geometry of the weighted $\ell_1$ ball
  $B_{\ell_1}[\bm{w},\delta]$, and more specifically its intersection
  with the non-negative hyperoctant of $\Real^l$. The reason for studying
  only the non-negative hyperoctant is justified by
  Lemma~\ref{lem:1st.hyperoctant}. Two points $\bm{x}_1, \bm{x}_2$ of
  $\Real^l$ are taken to demonstrate the concepts introduced in the
  manuscript. Notice that $P_{H_l^-}(\bm{x}_1)>\bm{0}$, which implies
  by Lemma~\ref{lem:proj.l1}.\ref{lem:strictly.in.hyperoctant} that
  $P_{Q_l}(\bm{x}_1) = P_{H_l^-}(\bm{x}_1)$. Notice also the case of
  $\bm{x}_2$ where some components of $P_{H_l^-}(\bm{x}_2)$ obtain
  negative values. Such a case mobilizes
  Lemma~\ref{lem:proj.l1}.\ref{lem:not.strictly.in.hyperoctant}.}
\label{fig:weighted.l1.ball}
\end{figure}

The following theorem computes, in a finite number of steps, the exact
projection of a point onto a weighted $\ell_1$ ball. The result
generalizes the projection mapping computed for the case of the
classical unweighted $\ell_1$ ball in \cite{Duchi08}. In words, the
projection mapping exploits the part of the weighted $\ell_1$ ball
that lies in the non-negative hyperoctant of the associated
space. This is because the projection of a point onto the weighted
$\ell_1$ ball lies always in the same hyperoctant as the point
itself. Hence, one may always choose to map the problem on the
non-negative hyperoctant, work there, and then return to the original
hyperoctant of the space, where the point lies. The part of the
weighted $\ell_1$ norm, that lies in the non-negative hyperoctant, can
be seen as the intersection of a closed halfspace and the non-negative
hyperoctant, see Fig.~\ref{fig:weighted.l1.ball}. It turns out that if
the projection of a point on this specific halfspace has all its
components positive, e.g., point $\bm{x}_1$ in
Fig.~\ref{fig:weighted.l1.ball}, then the projection on the halfspace
and the projection of the point on the weighted $\ell_1$ ball
coincide. If, however, some of the components of the projection onto
the halfspace are non-positive, e.g., point $\bm{x}_2$ in
Fig.~\ref{fig:weighted.l1.ball}, the corresponding dimensions are
ignored and the projection takes place in the resulting lower
dimensional space. It turns out that this projection coincides with
the projection of the point on the weighted $\ell_1$ ball. The
previous procedure is formally summarized next.

\begin{theorem}\label{thm:the.projection.onto.l1}
Given an $\bm{h}\in\Real^L\setminus B_{\ell_1}[\bm{w}_n,\delta]$, the
following recursion computes, in a finite number of steps (at most
$L$), the projection of $\bm{h}$ onto the ball
$B_{\ell_1}[\bm{w}_n,\delta]$, i.e., the (unique) vector
$P_{B_{\ell_1}[\bm{w}_n,\delta]}(\bm{h})\in\Real^L$. The case of
$\bm{h}\in B_{\ell_1}[\bm{w}_n,\delta]$ is trivial, since
$P_{B_{\ell_1}[\bm{w}_n,\delta]}(\bm{h})=\bm{h}$.

\begin{enumerate}
\item\label{algo:1st.hyperoctant} Form the vector
      $[|h_1|/w_{n,1},\ldots,|h_L|/w_{n,L}]^T\in\Real^L$.

\item\label{algo:sorting} Sort the previous vector in a non-ascending
  order (this takes $\mathcal{O}(L\log_2 L)$ computations), so that
  $[|h_{\tau(1)}|/w_{n,\tau(1)},\ldots,|h_{\tau(L)}|/w_{n,\tau(L)}]^T$,
  with $|h_{\tau(1)}|/w_{n,\tau(1)} \geq \cdots \geq
  |h_{\tau(L)}|/w_{n,\tau(L)}$, is obtained. The notation $\tau$
  stands for the permutation, which is implicitly defined by
  the sorting algorithm. Keep in memory the inverse $\tau^{-1}$ which
  moves the sorted elements back to the original positions.

\item Let $r_1 := L$.

\item Let $l=1$. While $l\leq L$, do the following.

\begin{enumerate}

\item Let $\lambda_* := l$.\label{start.loop}

\item\label{algo:active.dimensions} Find the maximum $j_*$ among those
  $j\in\overline{1,r_l}$ such that
  $\frac{|h_{\tau(j)}|}{w_{n,\tau(j)}} > \frac{\sum_{i=1}^{r_l}
  w_{n,\tau(i)} |h_{\tau(i)}| -\delta} {\sum_{i=1}^{r_l}
  w_{n,\tau(i)}^2}$.

\item\label{algo:found.the.projection} If $j_* = r_l$ then break the loop.

\item\label{algo:drop.dimensions} Otherwise, set $r_{l+1} := j_*$.

\item Increase $l$ by $1$, and go back to Step~\ref{start.loop}.

\end{enumerate}

\item Form the vector $\hat{\bm{p}} \in\Real^{r_{\lambda_*}}$ whose
      $j$-th component is given by $\hat{p}_j := |h_{\tau(j)}| -
  \frac{\sum_{i=1}^{r_{\lambda_*}} w_{n,\tau(i)}|h_{\tau(i)}| -\delta}
       {\sum_{i=1}^{r_{\lambda_*}} w_{n,\tau(i)}^2} w_{n,\tau(j)}$.

\item Use the inverse mapping $\tau^{-1}$, met in step 2, to insert the
      number $\hat{p}_j$ into the $\tau^{-1}(j)$ position of the
      $L$-dimensional vector $\bm{p}$, i.e., $p_{\tau^{-1}(j)} :=
      \hat{p}_j$, $\forall j\in\overline{1,r_{\lambda_*}}$, and fill in the
      remaining $L-r_{\lambda_*}$ positions of $\bm{p}$ with zeros.

\item The desired projection is
  $P_{B_{\ell_1}[\bm{w}_n,\delta]}(\bm{h}) = [\sign(h_1) p_1, \ldots,
  \sign(h_L) p_L]^T \in\Real^L$, where the symbol $\sign(\cdot)$
  stands for the sign of a real number.
\end{enumerate}\qed
\end{theorem}

\begin{proof}
The proof is given in Appendix~\ref{app:project.onto.l1.ball}. It
follows a geometric approach, instead of the Lagrange multipliers
methodology, which was followed in \cite{Duchi08} for the case of the
unweighted $\ell_1$ norm.
\end{proof}

\section{Weighting the $\ell_1$ Ball}\label{sec:weights.specs}

Motivated by the strategy adopted in \cite{CandesWakinBoyd08}, the
sequence of weights $(\bm{w}_n)_{n\in\Natural}$ is designed as
follows; let the $i$-th component of the vector $\bm{w}_n$ be given by
\begin{equation}
w_{n,i} := \frac{1}{|h_{n,i}| + \epsilon_n'},\ \forall
 i\in\overline{1,L},\ \forall n\in\Natural,\label{weights}
\end{equation}
where $(\epsilon'_n)_{n\in\Natural}$ is a sequence of small
positive parameters, which are used in order to avoid division by
zero. An illustration of the induced geometry can be seen in
Fig.~\ref{fig:projections}. A way to design the parameters
$(\epsilon'_n)_{n\in\Natural}$ will be given in the next
section. The corresponding algorithm will be referred to as the
Adaptive Projection Algorithm onto Weighted $\ell_1$ balls
(APWL1). The unweighted case, i.e., when $\bm{w}_n:=\bm{1}$, $\forall
n\in\Natural$, will be also considered and is denoted as APL1.

\begin{remark}
The radius $\delta$ of the $\ell_1$ norm, on which we project, depends
on whether the unweighted or the weighted version is adopted. In the
unweighted $\ell_1$ norm case, the optimum value of the radius is
apparently $\delta :=\norm{\bm{h}_*}_{\ell_1}$. However, in the
weighted case, $\delta$ is set equal to
$S=\norm{\bm{h}_*}_{\ell_0}$. The reason for this is the following.

Consider the desirable situation where our sequence of estimates
$(\bm{h}_n)_{n\in\Natural}$ converges to $\bm{h}_*$, i.e.,
$\lim_{n\rightarrow\infty}\bm{h}_n = \bm{h}_*$. Moreover, let
$\epsilon_n'\geq \epsilon'>0$, $\forall n\in\Natural$, where
$\epsilon'$ is a user-defined parameter. Then, $\sum_{i=1}^L w_{n,i}
|h_{n,i}| \leq \sum_{i=1}^L \frac{|h_{n,i}|}{|h_{n,i}| +
  \epsilon'}$, $\forall n\in\Natural$, and thus,
\begin{align*}
\limsup_{n\rightarrow\infty} \sum_{i=1}^L w_{n,i} |h_{n,i}| & \leq
\limsup_{n\rightarrow\infty} \sum_{i=1}^L \frac{|h_{n,i}|}{|h_{n,i}|
  + \epsilon'} = \lim_{n\rightarrow\infty} \sum_{i=1}^L
\frac{|h_{n,i}|}{|h_{n,i}| + \epsilon'} \\
& = \sum_{i\in\supp(\bm{h}_*)} \frac{|h_{*,i}|}{|h_{*,i}| + \epsilon'} +
\sum_{i\notin\supp(\bm{h}_*)} \frac{|h_{*,i}|}{|h_{*,i}| + \epsilon'}
< \sum_{i\in\supp(\bm{h}_*)} \frac{|h_{*,i}|}{|h_{*,i}|} =
\norm{\bm{h}_*}_{\ell_0}.
\end{align*}
The previous strict inequality and the definition of $\limsup$ suggest that
there exists an $m_1\in\Natural$ such that
$\forall n\geq m_1$ we have $\sum_{i=1}^L w_{n,i} |h_{n,i}| \leq
\norm{\bm{h}_*}_{\ell_0}$. In other words, we obtain that $\forall
n\geq m_1$, $\bm{h}_n \in B_{\ell_1}[\bm{w}_n,
  \norm{\bm{h}_*}_{\ell_0}]$. Hence, a natural choice for $\delta$ in
the design of the constraint set $B_{\ell_1}[\bm{w}_n,\delta]$ is
$\norm{\bm{h}_*}_{\ell_0}$. At least, such a choice is justified
$\forall n\geq m_1$, since it becomes a necessary condition for having
$(\bm{h}_n)_{n\in\Natural}$ converge to the desirable $\bm{h}_*$.\qed
\end{remark}

\section{Convergence Properties of the
  Algorithm}\label{sec:convergence}

It can be shown that, under certain assumptions, the previous algorithm
produces a sequence of estimates $(\bm{h}_n)_{n\in\Natural}$, which
converges to a point located arbitrarily close to an intersection as
in \eqref{the.problem}. The convergence of the algorithm is guaranteed
even if a finite number of closed convex sets do not share any
nonempty intersection with the rest of the convex constraints in
\eqref{the.problem}. This is important, since it allows for a finite
number of data outliers not to disturb the convergence of the
algorithm.

\noindent{\bf Assumptions.}

\begin{enumerate}

\item\label{ass:nonempty.intersection} Define $\forall n\in\Natural$,
  $\Omega_n := B_{\ell_1}[\bm{w}_n,\delta] \cap
  \left(\bigcap_{j\in\mathcal{J}_n} S_j[\epsilon]\right)$, i.e., the
  set $\Omega_n$ is defined as the intersection of the weighted
  $\ell_1$ ball and the hyperslabs that are considered at time
  $n$. Assume that there exists a $z_0\in\Natural$ such that $\Omega
  := \bigcap_{n\geq z_0} \Omega_n \neq\emptyset$. That is, with the
  exception of a finite number of $\Omega_n$s, the rest of them have a
  nonempty intersection.

\item\label{ass:lambda.epsilon} Choose a sufficiently small
  $\epsilon''>0$, and let $\forall n\in\Natural$,
  $\frac{\mu_n}{\mathcal{M}_n}\in [\epsilon'',2-\epsilon'']$.

\item\label{ass:int.Omega} The interior of $\Omega$ is nonempty, i.e.,
  $\interior(\Omega) \neq \emptyset$. For the definition of
  $\interior(\cdot)$ see Fact \ref{fact:special.Fejer} in Appendix
  \ref{app:analysis.algo}.

\item\label{ass:omega} Assume that $\check{\omega}:=
  \inf\{\omega_j^{(n)}:\ j\in\mathcal{J}_n, n\in\Natural\}>0$. In
  words, none of the weights, used to combine the projections onto the
  hyperslabs, will fade away as time $n$ advances.

\end{enumerate}\qed

\begin{theorem}[Convergence analysis of
the Algorithm]\label{thm:the.analysis}

Under the previously adopted assumptions, the following properties can
 be established.

\begin{enumerate}

\item Every update takes us closer to the intersection $\Omega$. In
  other words, the convergence is monotonic, that is, $\forall n\geq
  z_0$, $d(\bm{h}_{n+1},\Omega) \leq d(\bm{h}_n,\Omega)$.

\item\label{thm:distance.2.hyperslabs} Asymptotically, the distance of
      the obtained estimates from the respective hyperslabs tends to
      zero. That is, $\lim_{n\rightarrow\infty}
      \max\{d(\bm{h}_n,S_j[\epsilon]):\ j\in\mathcal{J}_n\}=0$.

\item\label{thm:distance.2.l1.balls} Similarly, the distance of the
  obtained estimates from the respective weighted $\ell_1$ balls tends
  asymptotically to zero. That is, $\lim_{n\rightarrow\infty}
  d(\bm{h}_n,B_{\ell_1}[\bm{w}_n,\delta])=0$.

\item\label{thm:liminf} Finally, there exists an $\tilde{\bm{h}}_*\in\Real^L$
  such that the sequence of estimates $(\bm{h}_n)_{n\in\Natural}$
  converges to, i.e., $\lim_{n\rightarrow\infty} \bm{h}_n =
      \tilde{\bm{h}}_*$, and that
\begin{equation*}
\tilde{\bm{h}}_* \in \left(\overline{\liminf_{n\rightarrow\infty}}
 B_{\ell_1}[\bm{w}_n,\delta]\right) \cap
\left(\overline{\liminf_{n\rightarrow\infty}}
 \bigcap_{j\in\mathcal{J}_n} S_j[\epsilon] \right).
\end{equation*}
Here, $\liminf_{n\rightarrow\infty} C_n:= \bigcup_{n\geq 0}
     \bigcap_{m\geq n} C_m$, for any sequence
     $(C_n)_{n\in\Natural}\subset \Real^L$, and the overline denotes the
     closure of a set. In other words, the algorithm converges to a
     point that lies arbitrarily close to an intersection of all the
      involved property sets.
\end{enumerate}\qed
\end{theorem}

\begin{proof}
The proof of these results, several auxiliary concepts, as well as
details on which assumptions are activated, in order to prove each
result, can be found in Appendix~\ref{app:analysis.algo}.
\end{proof}

\begin{remark}
Regarding Assumption~\ref{ass:int.Omega}, the condition
$\interior\bigcap_{n\in\Natural} B_{\ell_1}[\bm{w}_n,\delta]
\neq\emptyset$ can be easily satisfied. To see this, choose
arbitrarily a sufficiently small $\epsilon'>0$, and let in
\eqref{weights}: $\epsilon'_n \geq \epsilon'$, $\forall
n\in\Natural$. Then, notice by Fig.~\ref{fig:weighted.l1.ball}
that $\forall n\in\Natural$, $\forall i\in\overline{1,L}$,
$\frac{\delta}{w_{n,i}} = \delta(|h_{n,i}|+\epsilon_n') \geq \delta
\epsilon_n' \geq \delta\epsilon'$. This clearly implies that $\forall
n\in\Natural$, $B_{\ell_1}[\bm{1},\delta\epsilon'] \subset
B_{\ell_1}[\bm{w}_n,\delta]$, $\forall n\in\Natural$. It is easy now
to verify that $B(\bm{0},\frac{\delta\epsilon'}{\sqrt{L}}) :=
\{\bm{h}\in\Real^L:\ \norm{\bm{h}} < \frac{\delta\epsilon'}{\sqrt{L}}
\} \subset B_{\ell_1}[\bm{1},\delta\epsilon'] \subset
\bigcap_{n\in\Natural} B_{\ell_1}[\bm{w}_n,\delta]$, which implies, of
course, that $\bm{0}\in \interior\bigcap_{n\in\Natural}
B_{\ell_1}[\bm{w}_n,\delta]\neq\emptyset$.\qed
\end{remark}

\section{Performance evaluation}\label{sec:simulation.examples}

In this section, the performance of the proposed algorithms is
evaluated against both time-invariant and time-varying signals and
systems. It is also compared to a number of other online algorithms
such as the Zero-Attracting LMS (ZA-LMS) \cite{ChenHero09}, the
Reweighted ZA-LMS (RZA-LMS) \cite{ChenHero09}, and the
Recursive LASSO (RLASSO) \cite{Angelosante09}. Moreover, the LASSO
performance, when solved with batch methods
\cite{BergFriedlander:2008, spgl1:2007} is also given, since it serves
as a benchmark for the best achievable performance with
$\ell_1$-regularized LS solvers. All the performance curves are the
result from ensemble averaging of 100 independent runs.  Moreover, for
all the projection based algorithms tested, in all simulation
examples, $\mu_n$ was set equal to $\mathcal{M}_n/2$ and the
hyperslabs parameter $\epsilon$ was set equal to $1.3 \times \sigma$,
with $\sigma$ being the noise standard deviation. Even though such a
choice may not be necessarily optimal, the proposed algorithms turn
out to be relatively insensitive to the values of these parameters.
Finally, $\omega_j^{(n)}$ of \eqref{algo} are set equal to
$1/q$, $\forall j\in\mathcal{J}_n$, $\forall n\in\Natural$.

\subsection{Time-invariant case.}

\begin{figure*}[t]
\begin{minipage}[b]{1.0\linewidth}
  \centering
 \centerline{\epsfig{figure= 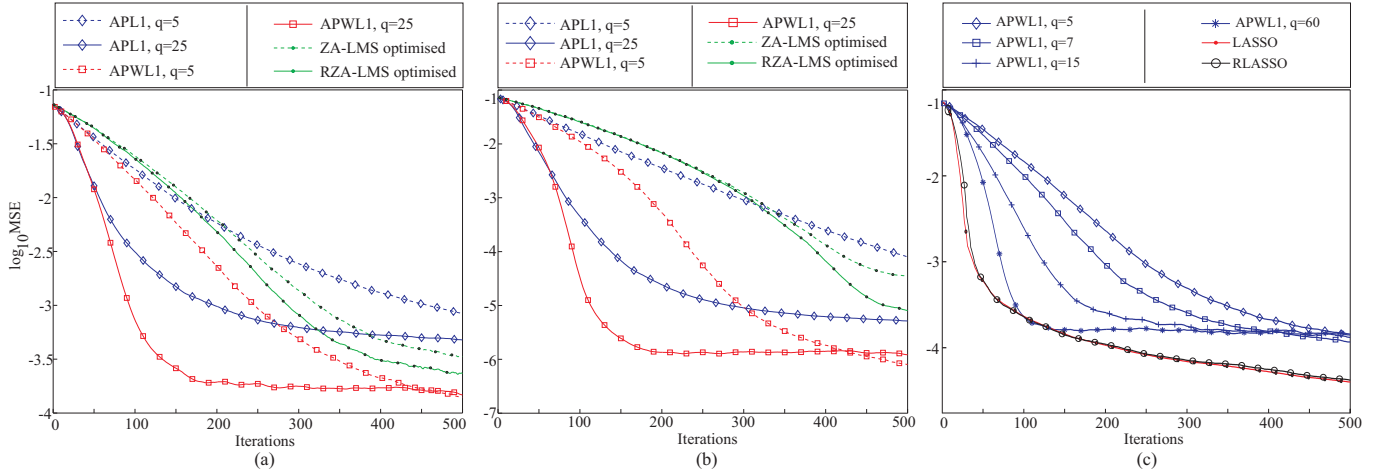,width=18cm}}
\end{minipage}
\caption{Sparse system identification example with $L=100$ and
 $S=5$. (a) and (b) shows the performance comparison of the proposed
 techniques with the LMS-based methods for high and low noise
 respectively. (c) shows the effect of different $q$ values in
 comparison to the LASSO performance.}  \label{fig:First_example}
\end{figure*}
In this simulation example, a time-invariant system having $L=100$
coefficients is used. The system is sparse with $S=5$, i.e., it has
only five nonzero coefficients, which are placed in arbitrary
positions. The input signal $\bm{x}$ is formed with entries drawn from
a zero-mean normal distribution with variance $1$.

In Figs.~\ref{fig:First_example}(a) and \ref{fig:First_example}(b) the
performance of the new algorithm is compared with that obtained by the
LMS-based methods, in different noise levels. The noise variance was
set equal to two different values, i.e., $\sigma_n^2=0.1$ and
$\sigma_n^2=0.001$ corresponding to SNR values of approximately $-3$dB
and $17$dB, respectively. Two different values of the parameter $q$
have been considered, namely 5 an 25. Moreover, with respect to the
ZA-LMS and the RZA-LMS, the ``optimized'' tag indicates that the free
parameters $\mu$ and $\rho$ were optimized, in order to give the best
performance at the 450th iteration. A different parameter setup could
lead to faster convergence of both LMS-based methods, albeit at the
expense of higher error-floors. In Fig.~\ref{fig:First_example}(a) we
observe that APWL1 exhibits the best performance both with respect to
convergence speed as well as steady-state error floor.  In fact, the
larger the value of $q$ is the faster the convergence
becomes. However, when the unweighted $\ell_1$ ball is used (APL1),
the method assumes relatively high error-floors, worse than both the
LMS-based methods.

In all the cases, unless the contrary is explicitly stated, the
adopted values for $\delta$ were: $\delta :=\norm{\bm{h}_*}_{\ell_1}$
and $\delta := S$ for the APL1 and the APWL1 respectively. The
sensitivity of these methods, on using different values of $\delta$,
will be discussed in section \ref{sec:sensitivitydelta}. Moreover, the
adaptation strategy of $\epsilon'_n$ in \eqref{weights} was decided
upon the following observation. A very small $\epsilon'_n$, in the
range of $[0.001,0.01]$, leads to low error-floors but the convergence
speed is compromised. On the other hand, when $\epsilon'_n$ is
relatively large, e.g., $\epsilon'_n \ge 0.1$, then fast convergence
speed is favored at the expense of a higher steady state error
floor. In order to tackle this issue efficiently, $\epsilon'_n$ can
start with a high value and then getting gradually smaller.
Although other scenarios may be possible, 
in all the time invariant examples, 
we have chosen: $\epsilon'_n:=\epsilon'+\frac{1}{n+1}$, $\forall
n\in\Natural$, where $\epsilon'$ is a user-defined small positive
constant.

Fig.~\ref{fig:First_example}(b) corresponds to a low noise level, where
the improved performance of the proposed algorithm, compared to that of
LMS-based algorithms, is even more enhanced.


It suffices to say, that this enhanced performance is achieved at the
expense of higher complexity. The LMS-based algorithms
require $O(L)$ multiply/add operations, while the APWL1 demands $q$
times more multiply/add operations. However, in a parallel processing
environment, the dependence on $q$ can be relaxed.

In Fig.~\ref{fig:First_example}(c) the performance improvement of
APWL1, as the $q$ value is increasing, is examined and compared to
performance of the RLASSO algorithm. In this system identification
case, the $L \times K$ regression matrices, $\bm{H}^{T}_{\tau}$, in
\cite{Angelosante09} are built using the input vectors
$(\bm{x}_n)_{n\in\Natural}$ according to $[\bm{x}_{L+\tau K-K},\ldots,
  \bm{x}_{L+\tau K-1}]$, for $\tau \in\Natural$, and
$K\in\Naturalstar$. Parameter $K$ was set equal to 5. As a reference,
the batch LASSO solution is also given, using the true delta value,
i.e., $\delta :=\norm{\bm{h}_*}_{\ell_1}$. The test is performed for
two different noise levels with the solid and the dotted performance
curves corresponding to $\sigma_n^2=0.1$ and
$\sigma_n^2=0.001$. Clearly, the convergence speed rapidly improves as
$q$ increases, and the rate of improvement is more noticeable in the
range of small values of $q$.


Observe that for large values of $q$, the 
performance gets ``closer'' 
to the one obtained by the LASSO and RLASSO
methods. Of course, the larger the $q$ the ``heavier'' the method
becomes from a computationally point of view. However, even for the
value of $q=60$ the complexity remains much lower than that of
RLASSO. The complexity of the latter algorithm rises up to the order of
$\mathcal{O}\left(rL^2\right)$, where $r$ is the number of iterations
for the cost function minimization in
\cite[(7)]{Angelosante09}. Indicatively, in the specific example $r$
needed to be larger than $L$ in order the method to converge for all the
realizations that were involved.

\begin{figure}[t]
\begin{minipage}[b]{1.0\linewidth}
  \centering
 \centerline{\epsfig{figure= 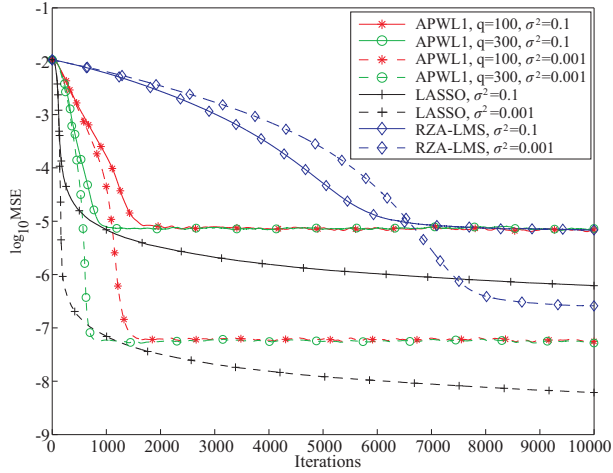,width=8cm}}
\end{minipage}
\caption{Sparse signal reconstruction example with $L=2000$ and $S=20$, for high and low noise levels.}
 \label{fig:examplelongsignal}
\end{figure}

In the sequel, we turn our attention to the estimation of a large
vector. We will realize it in the context of a signal reconstruction
task.  We assume a sparse signal vector of 2000 components with $S=20$
arbitrarily positioned nonzero components having values drawn from a
zero-mean normal distribution of unit variance.  In this case, the
observations are obtained from inner products of the unknown signal
with independent random measurement vectors, having values distributed
according to zero-mean normal distribution of unit variance. The
results, are depicted in Fig. \ref{fig:examplelongsignal} for
$\sigma_n^2=0.1$ (SNR=$-10$dB), and $\sigma_n^2=0.001$ (SNR=$10$dB), drawn
with solid and dashed lines, respectively. It is readily observed that
the same trend, which was discussed in the previous experiments, holds
true for this example. It must be pointed out that in the signal
reconstruction task, the input signal may not necessarily have the
shift invariance property \cite{SayedBook, Haykin}. Hence, techniques
that build around this property and have extensively been used in
order to reduce complexity in the context of LS algorithms, are not
applicable for such a problem. Both, LMS and the proposed algorithmic
scheme do not utilize this property.

\subsection{Time-varying case.}

\begin{figure}[t]
\begin{minipage}[b]{1.0\linewidth}
  \centering
 \centerline{\epsfig{figure= 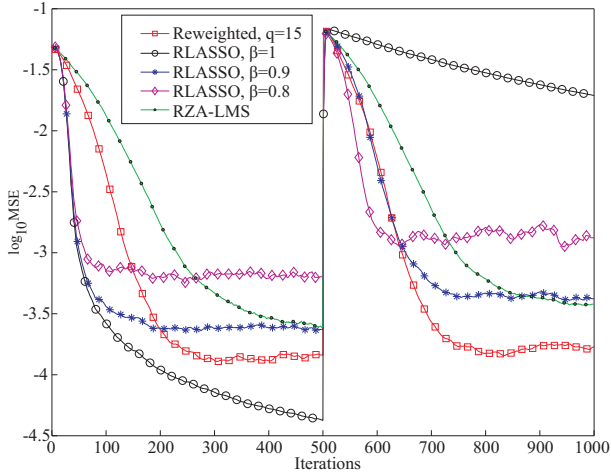,width=8cm}}
\end{minipage}
\caption{Time-varying sparse system identification example. The system
 impulse response changes abruptly at iteration $\#501$.}
 \label{fig:timevarying_example}
\end{figure}
It is by now well established in the adaptive filtering community,
e.g., \cite{SayedBook}, that convergence speed and tracking ability of an
algorithm do not, necessarily, follow the same trend. An algorithm may
have good converging properties, yet its tracking ability to time
variations may not be good, or vice versa. There are many cases where
LMS tracks better than the RLS. Although the theoretical analysis of
the tracking performance is much more difficult, due to the
non-stationarity of the environment, related simulated examples are
always needed to demonstrate the performance of an adaptive algorithm
in such environments. To this end, in this section, the performance of
the proposed method to track time-varying sparse systems is
investigated.  Both, the number of nonzero elements of $\bm{h}_*$ as
well as the values of the system's coefficients are allowed to undergo
sudden changes. This is a typical scenario used in adaptive filtering
in order to study the tracking performance of an algorithm in
practice. The system used in the experiments is 100 coefficients
long. The system change is realized as follows: For the first $500$
time instances, the first $5$ coefficients are set equal to 1. Then,
at time instance $n=501$ the second and the fourth coefficients are
changed to zero, and all the odd coefficients from $\#7$ to $\#15$ are
set equal to $1$. Note that the sparsity level, $S$, also changes at
time instance $n=501$, and it becomes $8$ instead of $5$. The results are
shown in Fig.~\ref{fig:timevarying_example} with the noise variance
being set equal to 0.1.

The curve indicated with squares corresponds to the proposed,
APWL1 method with $q=15$.  The performance of the RLASSO scheme
with forgetting factor $\beta=1$ is denoted by circles. The latter
clearly outperforms the rest of the methods up to time instance
500. This is expected, since LS-type of algorithms are known to have
fast converging properties. Note that up to this time instant, the
example coincides with that shown with solid curves in
Fig.~\ref{fig:First_example}(c). However, the algorithm lucks the
``agility'' of fast tracking the changes that take place after
convergence, due to its long memory.
In order to make it track faster,  the forgetting factor $\beta$
has to be decreased, in order to ``forget'' the remote past.
However, this affects its (initial) converging properties and in
particular the corresponding error floor.

When $\beta=0.8$ (curve denoted by diamonds), the tracking speed of the
RLASSO is significantly improved, albeit at the expense of
significantly increased error floor. The significant increase in the
error floor is also noticed in the first period, where it converges
fast, yet to a steady state of increased misadjustment error.
Adjusting the $\beta$ parameter to lead to lower error floors, one has
to sacrifice tracking speed. For the value of $\beta=0.9$, the RLASSO
(curve denoted by stars) achieves the same tracking speed as our
proposed method, however its error floor remains notably higher. In
fact, the steady state performance of RLASSO in this case, reaches the
levels of RZA-LMS (curve denoted by dots).

There are two issues related to the proposed method that have to be
discussed for the time-varying case. The first concerns the value of
$\delta$ and the other the adaptation strategy of $\epsilon'_n$.
Physical reasoning suggests that $\delta$, for the weighted $\ell_1$
ball, should be set equal to $5$ for the first $500$ iterations and then
take the value $8$. However, the actual sparsity levels can not be
known in advance. As a result, in the example of
Fig.~\ref{fig:timevarying_example}, $\delta$ was fixed to 9. As it
will be discussed soon, the method is rather insensitive against
overestimated $\delta$ values. Concerning $\epsilon'_n$, the
adaptation strategy discussed in the previous section, needs a slight
modification. Due to the fact that the system undergoes changes, the
algorithm has to be alert to track changes. 
In order to achieve this, the
algorithm has the ability to monitor abrupt changes of the orbit
$(\bm{h}_n)_{n\in\Natural}$. Whenever the estimated impulse response
changes considerably, and such a change also appears in the orbit
$(\bm{h}_n)_{n\in\Natural}$, $\epsilon'_n$ in \eqref{weights} is reset
to $\epsilon' + 1$ and it is gradually reduced similarly to the
previous example.

\section{Sensitivity of APWL1 to Non ideal Parameter
  Setting}\label{sec:sensitivity}

The robustness of any technique is affected by its sensitivity to non
``optimized'' configurations. In this section, the sensitivity of APWL1 on
$\delta$ and $\epsilon$ is examined. The sensitivity of APWL1 is
compared to the sensitivity that LASSO and LMS-based algorithms have
with respect to their associated parameters.


\subsection{Comparing to LASSO} \label{sec:sensitivitydelta}

\begin{figure}[t]
\begin{minipage}[b]{1.0\linewidth}
  \centering
 \centerline{\epsfig{figure= 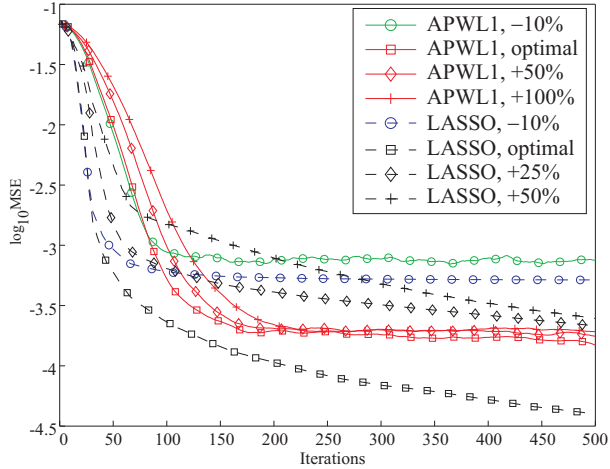,width=8cm}}
\end{minipage}
\caption{Sensitivity of APWL1 and LASSO to the $\delta$ parameter.}
 \label{fig:sensitivitywithlasso}
\end{figure}

In Fig.~\ref{fig:sensitivitywithlasso}, the solid lines indicated by
diamonds, crosses and circles correspond to the performance of the
APWL1, with $q=30$, when the true $\delta$ parameter is overestimated by
$50\%$, $100 \%$ or underestimated by $10\%$, respectively. The system
$\bm{h}_*$ under consideration has $L=100$, $S=5$ and
$\sigma_n^2=0.1$. The best performance, drawn with the solid curve
indicated with squares, is achieved when the APWL1 method is supplied with the
true $\delta$ value, i.e., when $\delta=S$. We observe that the
tolerance in $\delta$ underestimation is very limited, since even an
underestimation by $10 \%$ leads to a significant performance
degradation. On the other hand, APWL1 is rather insensitive to
overestimation. Indeed, overestimation even by $100\%$, compared to the
true value, leads to acceptable results. For comparison, the sensitivity
of the standard LASSO is presented with dashed lines. In this case, the
optimized $\delta$ value equals to $\norm{\bm{h}_*}_{\ell_1}$. The
sensitivity of LASSO is clearly higher, particularly to the steady-state
region. Observe, that only a $25\%$ deviation from the optimum value (dashed line with diamonds) causes enough performance degradation to
bring LASSO at a higher MSE regime, compared APWL1. Moreover, LASSO,
similarly to APWL1, exhibits limited tolerance in $\delta$
underestimated $\delta$ values.

\subsection{Comparing to LMS-based techniques.}
\label{sec:sensitivitylms}

\begin{figure}[t]
\begin{minipage}[b]{1.0\linewidth}
  \centering
 \centerline{\epsfig{figure= 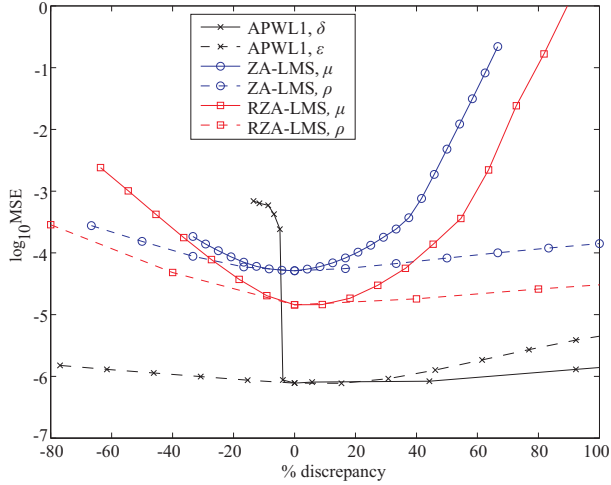,width=8cm}}
\end{minipage}
\caption{Sensitivity of the LMS-based methods on the $\mu$ and $\rho$ parameters compared to the sensitivity of APWL1 to $\delta$ and $\epsilon$.}
 \label{fig:sensitivitylms}
\end{figure}

Besides the $\delta$ parameter, APWL1 also needs specification of the
hyperslabs width, i.e., the parameter $\epsilon$. On the other hand,
LMS-based methods need the specification of $\mu$ and
$\rho$. Fig.~\ref{fig:sensitivitylms} shows the performance degradation
of APWL1 (curves with x-crosses), ZA-LMS (curves with circles) and
RZA-LMS (curves with rectangles), when they are configured with
parameter values which deviate from the ``optimum'' ones.  The x-axis
indicates deviation, from the ``optimal'' values, in percentage. The problem
setting is the one shown in Fig.~\ref{fig:First_example}(b) and the
reported MSE is always evaluated at time instance 450, where convergence
is assumed to have been achieved. The $0 \%$ discrepancy point,
coincides with the best achieved performance of each
method. For the LMS-based methods, the solid and dashed curves
correspond to $\mu$ and $\rho$, respectively. For the APWL1, the dashed
and the solid curves correspond to $\epsilon$ and $\delta$,
respectively. Starting with the latter parameter, as expected from the
discussion in section \ref{sec:sensitivitydelta}, even a slight
underestimation, i.e., negative deviation from the optimum, leads to a
sudden performance degradation. On the positive side, the method
exhibits a very low sensitivity. With respect to $\epsilon$, the
sensitivity of APWL1 is similar to the sensitivity exhibited by the
LMS-based methods on the $\rho$ parameter. However, LMS-based methods show
an increased sensitivity on the $\mu$ parameter for both negative and
positive deviation. In addition, the optimum $\mu$ value depends on the
length of $\bm{h}_*$, as it is the case with the standard non-regularized
LMS \cite{Haykin}.

\section{Conclusions}\label{sec:conclusions}

A novel efficient algorithm, of linear complexity, for sparse system
adaptive identification was presented, based on set theoretic
estimation arguments. Sparsity was exploited by the introduction of a
sequence of weighted $\ell_1$ balls. 
The algorithm consists of a sequence of projections on hyperslabs,
that measure data mismatch with respect to the training data, and on
weighted $\ell_1$ balls. The projection mapping on a weighted
$\ell_1$ ball has been derived and a full convergence proof of the
algorithm has been established. A comparative performance analysis,
using simulated data, was performed against the recently developed
online sparse LMS and sparse LS-type of algorithms.


\appendix
\section{The metric projection mapping onto the weighted $\ell_1$
  ball $B_{\ell_1}[\bm{w},\delta]$}\label{app:project.onto.l1.ball}

The results in this section are stated for any Euclidean space
$\Real^l$, where $l\in\overline{1,L}$. 
Moreover, given two vectors
$\bm{x}:=[x_1, \ldots, x_l]^T, \bm{y}:=[y_1, \ldots, y_l]^T
\in\Real^l$, then the notation $\bm{x}\leq(<) \bm{y}$ means that
$\forall i\in\overline{1,l}$, $x_i \leq(<) y_i$.


A well-known property of the metric projection mapping $P_C$ onto a closed
convex set $C$, which will be used in the sequel, is the
following \cite{BauschkeBorwein,StarkYangBook}:
\begin{equation}
\forall \bm{x}\in\Real^l, \forall
\bm{f}\in C, \quad \norm{\bm{x}-P_C(\bm{x})}^2 \leq \norm{\bm{x}-
\bm{f}}^2 - \norm{P_C(\bm{x})-\bm{f}}^2.\label{P.strongly.attracting}
\end{equation}

Define $Q_l:= B_{\ell_1}[\bm{w},\delta] \cap \mathbb{R}_{\geq 0}^l$,
where $\mathbb{R}_{\geq 0}^l$ stands for the non-negative
hyperoctant of $\Real^l$ (see Fig.~\ref{fig:weighted.l1.ball}). Define
also the following closed halfspace: $H_l^-:=
\{\bm{u}\in\Real^l:\ \sum_{i=1}^l w_i u_i = \bm{w}^T \bm{u} \leq
\delta\}$. Clearly the boundary of $H_l^-$ is the hyperplane: $H_l:=
\{\bm{u}\in\Real^l:\ \sum_{i=1}^l w_i u_i = \bm{w}^T \bm{u} =
\delta\}$. It is easy to verify that $Q_l = H_l^- \cap
\mathbb{R}_{\geq 0}^l$. Clearly, the boundary of $Q_l$ is $H_l \cap
\mathbb{R}_{\geq 0}^l$.

\begin{lemma}\label{lem:1st.hyperoctant}
\begin{enumerate}
\item\label{lem:same.sgn} For any $\bm{x}\in\Real^l$, the projection
  $P_{B_{\ell_1}[\bm{w},\delta]}(\bm{x})$ belongs to the same hyperoctant
  as $\bm{x}$ does, i.e., if $\bm{x}_* :=
  P_{B_{\ell_1}[\bm{w},\delta]}(\bm{x})$, then $\sign(x_{*,i}) =
  \sign(x_i)$, $\forall i\in\overline{1,l}$.

\item\label{lem:abs} Define the mapping $\abs: \bm{x}=
  [x_1,\ldots,x_l]^T \mapsto [|x_1|,\ldots,|x_l|]^T$,
    $\forall\bm{x}\in\Real^l$. It can be easily verified that $\abs$
    is an one-to-one mapping of any hyperoctant of $\Real^l$ onto
    $\mathbb{R}_{\geq 0}^l$, i.e., it is a bijection. Fix arbitrarily an
    $\bm{x}\in \Real^l$. Consider the mapping $\abs$ which bijectively
    maps the hyperoctant, in which $\bm{x}$ is located, to
    $\mathbb{R}_{\geq 0}^l$. Then,
     $P_{B_{\ell_1}[\bm{w},\delta]}(\bm{x})=\abs^{-1}
     \left(P_{B_{\ell_1}[\bm{w},\delta]}(\abs(\bm{x}))\right)$, where
     $\abs^{-1}$ stands for the inverse mapping of $\abs$. In other
words, in order to calculate the projection mapping onto
$B_{\ell_1}[\bm{w},\delta]$, it is sufficient to study only the case
of $\mathbb{R}_{\geq 0}^l$.
\end{enumerate}\qed
\end{lemma}

\begin{proof}
\begin{enumerate}
\item Without any loss of generality, assume that $\bm{x}$ belongs to
  the non-negative hyperoctant of $\Real^l$. We will show that also every
  component of $\bm{x}_*$ is, also, non-negative. In order to derive a
      contradiction, assume that there exist some
negative components of $\bm{x}_*$. To make the proof short, and with no
loss of generality, assume that the only negative component of
$\bm{x}_*$ is $x_{*,1}$. Define the vector $\bm{u}_*$ such that
$u_{*,1}:= 0$ and $u_{*,i}:= x_{*,i}$, $\forall
i\in\overline{2,l}$. Since $\bm{x}_* \in B_{\ell_1}[\bm{w},\delta]$, we
have that $\sum_{i=1}^l w_i |x_{*,i}|\leq \delta$, which easily leads to
$\sum_{i=1}^l w_i |u_{*,i}|= \sum_{i=2}^l w_i |x_{*,i}|\leq \sum_{i=1}^l
w_i |x_{*,i}|\leq \delta$, i.e., $\bm{u}_*\in
B_{\ell_1}[\bm{w},\delta]$. Moreover, notice that since
$x_{*,1}<0=u_{*,1}$, then $x_1 - x_{*,1} > x_1 - u_{*,1}= x_1\geq
0$. Hence, $\norm{\bm{x}-\bm{u}_*}^2 < (x_1 - x_{*,1})^2 + \sum_{i=2}^l
      (x_i - x_{*,i})^2 = \norm{\bm{x}-\bm{x}_*}^2$. This contradicts
      the fact that $\bm{x}_* =
P_{B_{\ell_1}[\bm{w},\delta]}(\bm{x})$, and establishes
Lemma~\ref{lem:1st.hyperoctant}.\ref{lem:same.sgn}.

\item Fix arbitrarily an $\bm{x}\in\Real^l$. As we have seen before,
  $P_{B_{\ell_1}[\bm{w},\delta]}(\bm{x})$ will be located in the same
  hyperoctant as $\bm{x}$. Let any $\bm{u}\in S_l$, where $S_l$ stands
  for the intersection of $B_{\ell_1}[\bm{w},\delta]$ with the same
  hyperoctant where $\bm{x}$ belongs to. As a result, we have that
  $\sign(x_i)=\sign(u_i)$, $\forall i\in\overline{1,l}$, and
\begin{align*}
\norm{\bm{x}- \bm{u}}^2 & = \sum_{i=1}^l (x_i - u_i)^2 = \sum_{i=1}^l
(\sign(x_i)|x_i| - \sign(u_i)|u_i|)^2\\
& = \sum_{i=1}^l (\sign(x_i)|x_i| - \sign(x_i)|u_i|)^2 = \sum_{i=1}^l
(|x_i| - |u_i|)^2\\
& = \norm{\abs(\bm{x}) - \abs(\bm{u})}^2.
\end{align*}
Notice here that $\abs$ is a
  bijection from $S_l$ to $Q_l$, so that the previous equality results
  into the following:
\begin{align*}
 &\norm{\abs(\bm{x}) - P_{B_{\ell_1}[\bm{w},\delta]}(\abs(\bm{x}))} =
\min_{\bm{u}'\in Q_l} \norm{\abs(\bm{x}) - \bm{u}'} = \min_{\bm{u}\in
 S_l} \norm{\abs(\bm{x}) - \abs(\bm{u})} \\
& = \min_{\bm{u}\in S_l}\norm{\bm{x}-\bm{u}} = \norm{\bm{x}-
 P_{B_{\ell_1}[\bm{w},\delta]}(\bm{x})}
 = \norm{\abs(\bm{x})-
 \abs\left(P_{B_{\ell_1}[\bm{w},\delta]}(\bm{x})\right)}.
\end{align*}
Therefore, by the uniqueness of the projection, $\abs\left(
      P_{B_{\ell_1}[\bm{w},\delta]}(\bm{x}) \right)=
      P_{B_{\ell_1}[\bm{w},\delta]}(\abs(\bm{x}))$, and
      Lemma~\ref{lem:1st.hyperoctant}.\ref{lem:abs} is established.
\end{enumerate}
\end{proof}

\begin{lemma}\label{lem:proj.l1}
Let an $\bm{x}\in \mathbb{R}_{\geq 0}^l\setminus Q_l$, and
\begin{equation}
\bm{x}_*:= P_{H_l^-}(\bm{x}) = \bm{x} -
\frac{\max\{0,\bm{x}^T\bm{w}-\delta\}}{\norm{\bm{w}}^2}
\bm{w}. \label{proj.halfspace}
\end{equation}

\begin{enumerate}

\item\label{lem:strictly.in.hyperoctant} Assume that
  $\bm{x}_* > \bm{0}$. Then, $P_{Q_l}(\bm{x}) = P_{H_l^-}(\bm{x})$.

\item\label{lem:not.strictly.in.hyperoctant} Make the following
  partitions $\bm{x} = \left[\begin{smallmatrix}
\hat{\bm{x}}\\ \tilde{\bm{x}}
\end{smallmatrix}\right]$, $\bm{x}_* = \left[\begin{smallmatrix}
\hat{\bm{x}}_*\\ \tilde{\bm{x}}_*
\end{smallmatrix}\right]$, where $\hat{l},
  \tilde{l}\in\overline{1,l}$, $\hat{l} + \tilde{l}=l$, and
  $\hat{\bm{x}}, \hat{\bm{x}}_* \in\Real^{\hat{l}}$, $\tilde{\bm{x}},
  \tilde{\bm{x}}_* \in\Real^{\tilde{l}}$. Assume, now, that there
  exists an $\tilde{l}\in\overline{1,l}$ such that
  $\tilde{\bm{x}}_*\leq \bm{0}$. Then,
\begin{equation*}
P_{Q_l}(\bm{x})^T=[P_{Q_{\hat{l}}}(\hat{\bm{x}})^T,~\bm{0}^T]^T
\end{equation*}

\end{enumerate}\qed
\end{lemma}

\begin{proof}
\begin{enumerate}

\item Since $\bm{x}_* := P_{H_l^-}(\bm{x})>\bm{0}$, it is clear that
$\bm{x}_* \in H_l^-\cap \mathbb{R}_{\geq 0}^l = Q_l$. Hence,
\begin{equation*}
\min_{\bm{u}\in Q_l}\norm{\bm{x}-\bm{u}}  \leq \norm{\bm{x} -
  \bm{x}_*}=  \norm{\bm{x} - P_{H_l^-}(\bm{x})}
 = \min_{\bm{u}\in H_l^-}\norm{\bm{x} - \bm{u}}
\leq \min_{\bm{u}\in Q_l}\norm{\bm{x}-\bm{u}},
\end{equation*}
where the last inequality comes from $Q_l \subset H_l^-$. Thus,
$\norm{\bm{x} - P_{H_l^-}(\bm{x})} = \min_{\bm{u}\in
  Q_l}\norm{\bm{x}-\bm{u}}$. Hence, by the uniqueness of the
projection, $P_{Q_l}(\bm{x})= P_{H_l^-}(\bm{x})$, and
Lemma~\ref{lem:proj.l1}.\ref{lem:strictly.in.hyperoctant} is
established.

\item Since $H_l$ is a hyperplane, $\forall \bm{u}\in H_l$,
$(\bm{u}-\bm{x}_*)^T (\bm{x}-\bm{x}_*)=0$, which implies, of course,
      that $\forall \bm{u}\in H_l\cap\mathbb{R}_{\geq 0}^l$,
      $(\bm{u}-\bm{x}_*)^T (\bm{x}-\bm{x}_*)=0$. Thus, $\forall
      \bm{u}\in H_l\cap\mathbb{R}_{\geq 0}^l$,
\begin{equation}
\norm{\bm{u}-\bm{x}}_{\Real^l}^2  = \norm{\bm{u}-\bm{x}_*}_{\Real^l}^2
+ \norm{\bm{x}_*-\bm{x}}_{\Real^l}^2
 = \norm{\hat{\bm{u}}- \hat{\bm{x}}_*}_{\Real^{\hat{l}}}^2 +
\norm{\tilde{\bm{u}}- \tilde{\bm{x}}_*}_{\Real^{\tilde{l}}}^2 +
\norm{\bm{x}_*-\bm{x}}_{\Real^l}^2.\label{Pythagoras}
\end{equation}
This in turn implies that
\begin{align}
P_{Q_l}(\bm{x}) & =
\argmin\{\norm{\bm{u}-\bm{x}}_{\Real^l}^2:\ \bm{u}\in
H_l\cap\mathbb{R}_{\geq 0}^l\} & \nonumber \\
& = \argmin\{\norm{\bm{u}-\bm{x}}_{\Real^l}^2:\ \hat{\bm{u}}\in
\Real^{\hat{l}}_{\geq 0}, \tilde{\bm{u}}\in \Real^{\tilde{l}}_{\geq
  0}, \hat{\bm{u}}^T \hat{\bm{w}} +
\tilde{\bm{u}}^T\hat{\bm{w}}=\delta \}\nonumber \\
& = \argmin\{ \norm{\hat{\bm{u}}-
  \hat{\bm{x}}_*}_{\Real^{\hat{l}}}^2 + \norm{\tilde{\bm{u}}-
  \tilde{\bm{x}}_*}_{\Real^{\tilde{l}}}^2:\ \hat{\bm{u}}\in
\Real^{\hat{l}}_{\geq 0}, \tilde{\bm{u}}\in \Real^{\tilde{l}}_{\geq
  0}, \hat{\bm{u}}^T \hat{\bm{w}} +
\tilde{\bm{u}}^T\hat{\bm{w}}=\delta\}.
 \label{equiv.argmin}
\end{align}

By our initial assumption $\tilde{\bm{x}}_*\leq \bm{0}$. Hence, it is
easy to verify that $\forall \hat{\bm{u}}\in
\Real^{\hat{l}}_{\geq 0}$, $\forall \tilde{\bm{u}}\in
\Real^{\tilde{l}}_{\geq 0}\setminus \{\bm{0}\}$, $\norm{\hat{\bm{u}}-
  \hat{\bm{x}}_*}_{\Real^{\hat{l}}}^2+ \norm{\bm{0}-
  \tilde{\bm{x}}_*}_{\Real^{\tilde{l}}}^2 < \norm{\hat{\bm{u}}-
  \hat{\bm{x}}_*}_{\Real^{\hat{l}}}^2+ \norm{\tilde{\bm{u}}-
  \tilde{\bm{x}}_*}_{\Real^{\tilde{l}}}^2$, which evidently suggests
      that
\begin{align}
& \argmin\{ \norm{\hat{\bm{u}}-
  \hat{\bm{x}}_*}_{\Real^{\hat{l}}}^2 + \norm{\tilde{\bm{u}}-
  \tilde{\bm{x}}_*}_{\Real^{\tilde{l}}}^2:\ \hat{\bm{u}}\in
\Real^{\hat{l}}_{\geq 0}, \tilde{\bm{u}}\in \Real^{\tilde{l}}_{\geq
  0}, \hat{\bm{u}}^T \hat{\bm{w}} +
\tilde{\bm{u}}^T\hat{\bm{w}}=\delta\}\nonumber \\
&\hspace{10pt} = \argmin\{ \norm{\hat{\bm{u}}-
  \hat{\bm{x}}_*}_{\Real^{\hat{l}}}^2 + \norm{\tilde{\bm{u}}-
  \tilde{\bm{x}}_*}_{\Real^{\tilde{l}}}^2:\ \hat{\bm{u}}\in
\Real^{\hat{l}}_{\geq 0}, \hat{\bm{u}}^T
\hat{\bm{w}} =\delta, \tilde{\bm{u}}=\bm{0}\}
\label{special.argmin}
\end{align}

Now, since $\bm{x}\in \mathbb{R}_{\geq 0}^l\setminus Q_l$, it is clear
  by the geometry of the $Q_l$ that $P_{Q_l}(\bm{x})$ will be located on
  $H_l\cap\mathbb{R}_{\geq 0}^l$. Hence, by \eqref{Pythagoras},
  \eqref{equiv.argmin}, and \eqref{special.argmin}, one can verify the
  following:
\begin{align*}
P_{Q_l}(\bm{x}) & =
\argmin\{\norm{\bm{u}-\bm{x}}_{\Real^l}^2:\ \bm{u}\in H_l\cap\mathbb{R}_{\geq 0}^l\}
\\
& = \argmin\{ \norm{\hat{\bm{u}}-
  \hat{\bm{x}}_*}_{\Real^{\hat{l}}}^2 + \norm{\tilde{\bm{u}}-
  \tilde{\bm{x}}_*}_{\Real^{\tilde{l}}}^2:\ \hat{\bm{u}}\in
\Real^{\hat{l}}_{\geq 0}, \tilde{\bm{u}}\in \Real^{\tilde{l}}_{\geq
  0}, \hat{\bm{u}}^T \hat{\bm{w}} +
\tilde{\bm{u}}^T\hat{\bm{w}}=\delta\} \\
& = \argmin\{ \norm{\hat{\bm{u}}-
  \hat{\bm{x}}_*}_{\Real^{\hat{l}}}^2 + \norm{\tilde{\bm{u}}-
  \tilde{\bm{x}}_*}_{\Real^{\tilde{l}}}^2:\ \hat{\bm{u}}\in
\Real^{\hat{l}}_{\geq 0}, \hat{\bm{u}}^T
\hat{\bm{w}} =\delta, \tilde{\bm{u}}=\bm{0}\}\\
& = \argmin\{\norm{\bm{u}-\bm{x}}_{\Real^l}^2:\ \hat{\bm{u}}\in
\Real^{\hat{l}}_{\geq 0}, \hat{\bm{u}}^T
\hat{\bm{w}} =\delta, \tilde{\bm{u}}=\bm{0}\}\\
& = \argmin\{\norm{\hat{\bm{u}}-\hat{\bm{x}}}_{\Real^{\hat{l}}}^2 +
\norm{\tilde{\bm{u}}-\tilde{\bm{x}}}_{\Real^{\tilde{l}}}^2:\ \hat{\bm{u}}\in
\Real^{\hat{l}}_{\geq 0}, \hat{\bm{u}}^T
\hat{\bm{w}} =\delta, \tilde{\bm{u}}=\bm{0}\}\\
& =
\argmin\{\norm{\hat{\bm{u}}-\hat{\bm{x}}}_{\Real^{\hat{l}}}^2:\ \hat{\bm{u}}
\in \Real^{\hat{l}}_{\geq 0}, \hat{\bm{u}}^T \hat{\bm{w}} =\delta,
\tilde{\bm{u}}=\bm{0}\} \\
&
=\argmin\{\norm{\hat{\bm{u}}-\hat{\bm{x}}}_{\Real^{\hat{l}}}^2:\ \hat{\bm{u}}
\in H_{\hat{l}}\cap\mathbb{R}_{\geq 0}^{\hat{l}}, \tilde{\bm{u}}=\bm{0}\}\\
& = \begin{bmatrix}
P_{Q_{\hat{l}}}(\hat{\bm{x}})\\ \bm{0}
\end{bmatrix}.
\end{align*}
This establishes
Lemma~\ref{lem:proj.l1}.\ref{lem:not.strictly.in.hyperoctant}.
\end{enumerate}
\end{proof}

\begin{lemma}\label{lem:sorting.makes.things.easy}
Assume an $\bm{x}\in\mathbb{R}_{\geq 0}^l$ such that $\forall
i\in\overline{1,l-1}$, $\frac{x_i}{w_i}\geq
\frac{x_{i+1}}{w_{i+1}}$. Moreover, let $\bm{x}_*:=
P_{H_l^-}(\bm{x})$. Assume that there exists an $i_0\in\overline{1,l}$
such that $x_{*,i_0}\leq 0$. Then, $\forall i\geq i_0$, $x_{*,i}\leq
0$.\qed
\end{lemma}

\begin{proof}
Here we consider only the case where $\bm{x}\in\mathbb{R}_{\geq
  0}^l\setminus Q_l$, i.e., $\bm{x}^T \bm{w} -\delta >0$. Notice by
\eqref{proj.halfspace} that
\begin{equation}
x_{*,i}\leq 0 \Leftrightarrow \frac{x_i}{w_i}\leq
\frac{\bm{x}^T\bm{w}-\delta}{\norm{\bm{w}}^2}. \label{equiv.leq.0}
\end{equation}
Now, notice also that by the construction of $\bm{x}$ and by our initial
assumption, we have that
\begin{equation*}
\forall i\geq i_0,\quad \frac{x_i}{w_i}\leq \frac{x_{i_0}}{w_{i_0}}
\leq \frac{\bm{x}^T\bm{w}-\delta}{\norm{\bm{w}}^2}.
\end{equation*}
However, by \eqref{equiv.leq.0}, this is equivalent to $x_{*,i}\leq 0$,
$\forall i\geq i_0$, which establishes
Lemma~\ref{lem:sorting.makes.things.easy}.
\end{proof}

\subsection{The proof of Theorem~\ref{thm:the.projection.onto.l1}.}

Notice that Step~\ref{algo:1st.hyperoctant} is due to
Lemma~\ref{lem:1st.hyperoctant}. Step~\ref{algo:active.dimensions} refers
to the attempt of the algorithm to locate the negative components of a
vector, according to
Lemma~\ref{lem:sorting.makes.things.easy}. Step~\ref{algo:found.the.projection}
refers to Lemma~\ref{lem:proj.l1}.\ref{lem:strictly.in.hyperoctant},
while Step~\ref{algo:drop.dimensions} corresponds to
Lemma~\ref{lem:proj.l1}.\ref{lem:not.strictly.in.hyperoctant}.

\section{The proof of
  Theorem~\ref{thm:the.analysis}} \label{app:analysis.algo}

\subsection{Preliminaries.}

\begin{definition}[Subgradient and subdifferential
    \cite{Hiriart}] \label{def:subgradient} Given a convex function
  $\Theta: \Real^L\rightarrow\Real$, a subgradient $\Theta'(\bm{x})$
  of $\Theta$ at $\bm{x}\in\Real^L$ is an element of $\Real^L$, which
  satisfies the following property: $\Theta'(\bm{x})^T (\bm{y}-\bm{x})
  + \Theta(\bm{x}) \leq \Theta(\bm{y})$, $\forall
  \bm{y}\in\Real^L$. The set of all the subgradients of $\Theta$ at
  the point $\bm{x}$ will be called the subdifferential of $\Theta$ at
  $\bm{x}$, and will be denoted by $\partial\Theta(\bm{x})$. Notice
  that if $\Theta$ is (G\^{a}teaux) differentiable at $\bm{x}$, then
  the only subgradient of $\Theta$ at $\bm{x}$ is its differential.\qed
\end{definition}

\begin{fact}\label{fact:partial.distance}
The subdifferential of the metric distance function $d(\cdot,C)$ to a
closed convex set $C\subset\Real^L$ is given as follows
\cite{Hiriart}:
\begin{equation*}
\partial d(\bm{x},C) = \begin{cases}
N_C(\bm{x}) \cap B[\bm{0},1], & \bm{x}\in C,\\
\frac{\bm{x}-P_C(\bm{x})}{d(\bm{x},C)}, & \bm{x}\notin C,
\end{cases}
\end{equation*}
where $N_C(\bm{x}):=\{\bm{y}\in\Real^L:\ \bm{y}^T(\bm{f}-\bm{x})\leq 0,
 \forall \bm{f} \in C\}$, and $B[\bm{0},1] := \{\bm{y}\in\Real^L:\
 \norm{\bm{y}}\leq 1\}$. Notice that $\forall \bm{x}\in\Real^L$,
 $\norm{d'(\bm{x},C)}\leq 1$, where $d'(\bm{x},C)$ stands for any
 subgradient in $\partial d(\bm{x},C)$.\qed
\end{fact}

We will give, now, an equivalent description of the Algorithm in
\eqref{algo}, which will help us in proving several properties of the
algorithm.

\begin{lemma}[Equivalent description of the
    Algorithm in \eqref{algo}]\label{lem:equivalent.description}
 Define the following non-negative functions:
\begin{equation}
\forall n\in\Natural, \forall\bm{x}\in\Real^L, \quad \Theta_n(\bm{x})
:= \begin{cases}
 \sum_{j\in\mathcal{J}_n} \frac{\omega_j^{(n)}
d(\bm{h}_n,S_j[\epsilon])}{L_n} d(\bm{x},S_j[\epsilon]), &
\text{if}\ \mathcal{I}_n \neq \emptyset,\\
0, & \text{if}\ \mathcal{I}_n = \emptyset,
\end{cases}\label{special.Theta_n}
\end{equation}
where $\mathcal{I}_n:= \{j\in\mathcal{J}_n:\ \bm{h}_n \notin
S_j[\epsilon]\}$, and $L_n := \sum_{j\in\mathcal{J}_n} \omega_j^{(n)}
d(\bm{h}_n,S_j[\epsilon])$. Then, \eqref{algo} can be
equivalently written as follows:
\begin{equation}
\forall n\in\Natural, \quad \bm{h}_{n+1} := \begin{cases}
P_{B_{\ell_1}[\bm{w}_n,\delta]}\left(\bm{h}_n - \lambda_n
\frac{\Theta_n(\bm{h}_n)}{\norm{\Theta_n'(\bm{h}_n)}^2}
\Theta_n'(\bm{h}_n) \right), & \text{if}\ \Theta_n'(\bm{h}_n) \neq
0,\\
P_{B_{\ell_1}[\bm{w}_n,\delta]}(\bm{h}_n), & \text{if}\
\Theta_n'(\bm{h}_n) = 0,
\end{cases} \label{basic.recursion}
\end{equation}
where $\lambda_n\in(0,2)$, $\forall n\in\Natural$, and
$\Theta'_n(\bm{h}_n)$ is any subgradient of $\Theta_n$ at
$\bm{h}_n$.\qed
\end{lemma}

\begin{proof}
First, a few comments regarding $L_n$ in \eqref{special.Theta_n} are
in order. It can be easily verified by the definition of
$\mathcal{I}_n$ that $\exists j_0\in \mathcal{J}_n:\ \bm{h}_n \notin
S_{j_0}[\epsilon]$, which is in turn equivalent to
$d(\bm{h}_n,S_{j_0}[\epsilon])>0$. Hence, $L_n\geq \omega_{j_0}^{(n)}
d(\bm{h}_n,S_{j_0}[\epsilon])>0$, and \eqref{special.Theta_n} is
well-defined. The reason for introducing $L_n$ in the design is to
give the freedom to the extrapolation parameter $\mu_n$ in
\eqref{algo} to be able to take values greater than or equal to $2$;
recall that $\mu_n\in (0,2\mathcal{M}_n)$ and $\mathcal{M}_n \geq 1$,
$\forall n\in\Natural$, in \eqref{Mn}.

Basic calculus on subdifferentials \cite{Hiriart} and the definition
of $\mathcal{I}_n$ suggest that
\begin{equation*}
\partial\Theta_n(\bm{x})
:= \begin{cases}
 \sum_{j\in\mathcal{I}_n} \frac{\omega_j^{(n)}
d(\bm{h}_n,S_j[\epsilon])}{L_n} \partial d(\bm{x},S_j[\epsilon]), &
\text{if}\ \mathcal{I}_n \neq \emptyset,\\
\{\bm{0}\}, & \text{if}\ \mathcal{I}_n = \emptyset.
\end{cases}
\end{equation*}
Hence, in the case where $\mathcal{I}_n\neq \emptyset$,
Fact~\ref{fact:partial.distance} implies that
\begin{align}
\Theta_n'(\bm{h}_n)  & = \sum_{j\in\mathcal{I}_n} \frac{\omega_j^{(n)}
d(\bm{h}_n,S_j[\epsilon])}{L_n} \frac{\bm{h}_n - P_{S_j[\epsilon]}
  (\bm{h}_n)}{d(\bm{h}_n,S_j[\epsilon])} \nonumber\\
& = \frac{1}{L_n} \sum_{j\in\mathcal{I}_n} \omega_j^{(n)} (\bm{h}_n -
P_{S_j[\epsilon]} (\bm{h}_n))\nonumber \\
 & = \frac{1}{L_n} \sum_{j\in\mathcal{J}_n} \omega_j^{(n)} (\bm{h}_n -
P_{S_j[\epsilon]} (\bm{h}_n)).\label{the.subgrad.Theta.n}
\end{align}
Clearly, if $\mathcal{I}_n\neq \emptyset$, then $\Theta_n'(\bm{h}_n) =
\bm{0}\Leftrightarrow \sum_{j\in\mathcal{J}_n}
\omega_j^{(n)}P_{S_j[\epsilon]} (\bm{h}_n) = \bm{h}_n$. Notice that
the same equivalence holds true also in the case where $\mathcal{I}_n
= \emptyset$, since in such a case $\bm{h}_n \in
\bigcap_{j\in\mathcal{J}_n} S_j[\epsilon] \Leftrightarrow \bm{h}_n =
P_{S_j[\epsilon]}(\bm{h}_n), \forall j\in\mathcal{J}_n$. In other
words, we have derived the following: $\forall n\in\Natural,
\Theta_n'(\bm{h}_n) = \bm{0}\Leftrightarrow \sum_{j\in\mathcal{J}_n}
\omega_j^{(n)}P_{S_j[\epsilon]} (\bm{h}_n) = \bm{h}_n$. By this
result, if we substitute \eqref{the.subgrad.Theta.n} in
\eqref{basic.recursion}, and if we define $\mu_n:= \lambda_n
\mathcal{M}_n$, $\forall n\in\Natural$, where $\mathcal{M}_n$ is given
in \eqref{Mn}, then we obtain the recursion given in \eqref{algo}.
\end{proof}

Next are a few observations on the function $\Theta_n$, which will help
 us to establish several convergence properties of the Algorithm in
 \eqref{algo}. First, notice that
\begin{align*}
\mathcal{I}_n  = \emptyset \Leftrightarrow \bm{h}_n \in
 \bigcap_{j\in\mathcal{J}_n} S_j[\epsilon] & \Leftrightarrow
 \left(\omega_j^{(n)}\bm{h}_n = \omega_j^{(n)} P_{S_j[\epsilon]}(\bm{h}_n),\
 \forall j\in\mathcal{J}_n \right)\\
& \Rightarrow \bm{h}_n = \sum_{j\in\mathcal{J}_n}
 \omega_j^{(n)}P_{S_j[\epsilon]}(\bm{h}_n) \Leftrightarrow
 \Theta'_n(\bm{h}_n) = \bm{0}.
\end{align*}
In the previous relation, the symbol $\Rightarrow$ becomes
$\Leftrightarrow$, if we assume that $\bigcap_{j\in\mathcal{J}_n}
S_j[\epsilon] \neq \emptyset$ \cite[Proposition
  2.12]{BauschkeBorwein}. Hence, if $\bigcap_{j\in\mathcal{J}_n}
S_j[\epsilon] \neq \emptyset$, then, $\mathcal{I}_n = \emptyset
\Leftrightarrow \bm{h}_n = \sum_{j\in\mathcal{J}_n}
\omega_j^{(n)}P_{S_j[\epsilon]}(\bm{h}_n) \Leftrightarrow
\Theta'_n(\bm{h}_n) = \bm{0}$. Moreover, in the case where
$\bigcap_{j\in\mathcal{J}_n} S_j[\epsilon] \neq \emptyset$, one can
verify also by the definition of $\Theta_n$ that
\begin{equation*}
\lev\Theta_n = \begin{cases}
\bigcap_{j\in\mathcal{I}_n} S_j[\epsilon], & \mathcal{I}_n \neq
		\emptyset,\\
\Real^L, & \mathcal{I}_n =\emptyset,
\end{cases}
\end{equation*}
where $\lev\Theta_n:= \{\bm{y}\in\Real^L:\ \Theta_n(\bm{y})\leq
0\}$.

Additionally, in the case where $\bigcap_{j\in\mathcal{J}_n}
 S_j[\epsilon] \neq \emptyset$, then we can establish the following
 equivalency: $\bm{h}_n \in \lev\Theta_n \Leftrightarrow \mathcal{I}_n =
 \emptyset$. This can be proved as follows. For the ``$\Leftarrow$''
 direction, we have that $\mathcal{I}_n = \emptyset \Leftrightarrow
 \bm{h}_n \in \bigcap_{j\in\mathcal{J}_n} S_j[\epsilon] \subset \Real^L
 = \lev\Theta_n$. As for the ``$\Rightarrow$'' direction, assume for a
 contradiction that $\mathcal{I}_n \neq \emptyset$. Then, by the
 preceding discussion, we have $\bm{h}_n \in \bigcap_{j\in\mathcal{I}_n}
 S_j[\epsilon]$, which is an absurd result if we recall the definition of
 $\mathcal{I}_n$. Thus, necessarily, $\mathcal{I}_n = \emptyset$, and
 the claim is proved. In other words, in the case where
 $\bigcap_{j\in\mathcal{J}_n} S_j[\epsilon] \neq \emptyset$, then,
 $\mathcal{I}_n = \emptyset \Leftrightarrow \Theta'_n(\bm{h}_n)=\bm{0}$,
 and thus
\begin{equation}
\bm{h}_n \in \lev\Theta_n \Leftrightarrow
 \Theta'_n(\bm{h}_n)=\bm{0}. \label{condition.4.lev.Theta.n}
\end{equation}

\begin{definition}[Subgradient projection
    mapping
    \cite{BauschkeCombettesWeak2Strong}] \label{def:subgrad.projection}
  Given a convex function $\Theta: \Real^L \rightarrow \Real$, such
  that $\lev\Theta\neq \emptyset$, define the \textit{subgradient
    projection mapping $T_{\Theta}: \Real^L\rightarrow\Real^L$ with
    respect to $\Theta$} as follows:
\begin{equation*}
T_{\Theta}(\bm{x}) := \begin{cases}
\bm{x} - \frac{\Theta(\bm{x})}{\norm{\Theta'(\bm{x})}^2}
\Theta'(\bm{x}), & \text{if}\ \bm{x}\notin \lev\Theta,\\
\bm{x}, & \text{if}\ \bm{x}\in\lev\Theta,
\end{cases}
\end{equation*}
where $\Theta'(\bm{x})$ stands for an arbitrarily fixed subgradient of
  $\Theta$ at $\bm{x}$. If $I$ stands for the identity mapping in
  $\Real^L$, the mapping $T_{\Theta}^{(\lambda)}:= I +
  \lambda(T_{\Theta}- I)$, $\lambda\in(0,2)$, will be called the {\em
  relaxed subgradient projection mapping}. Moreover, similarly to
  \eqref{P.strongly.attracting}, an important property of
  $T_{\Theta}^{(\lambda)}$ is the following
  \cite{BauschkeCombettesWeak2Strong}:
\begin{equation}
\forall \bm{x}\in\Real^L, \forall
\bm{f}\in\lev\Theta, \quad \frac{2-\lambda}{\lambda} \norm{\bm{x}-
T_{\Theta}^{(\lambda)}(\bm{x})}^2 \leq \norm{\bm{x}-
\bm{f}}^2 -
\norm{T_{\Theta}^{(\lambda)}(\bm{x})-\bm{f}}^2.
\label{T.strongly.attracting}
\end{equation}\qed
\end{definition}

Now, \eqref{P.strongly.attracting} and \eqref{T.strongly.attracting}
can be combined as follows.

\begin{lemma}\label{lem:compose.P.and.T} Let a closed convex set
 $C\subset\Real^L$, and a convex function $\Theta: \Real^L
 \rightarrow\Real$ such that $C\cap \lev\Theta \neq\emptyset$. Then,
\begin{equation*}
\forall \bm{x}\in\Real^L, \forall
\bm{f}\in C\cap\lev\Theta, \quad \frac{2-\lambda}{2} \norm{\bm{x}-
P_CT_{\Theta}^{(\lambda)}(\bm{x})}^2 \leq \norm{\bm{x}-
\bm{f}}^2 -
\norm{P_CT_{\Theta}^{(\lambda)}(\bm{x})-\bm{f}}^2.
\end{equation*}\qed
\end{lemma}

\begin{proof}
This is a direct consequence of \cite[Proposition
  1]{YamadaOguraAPSMNFAO}.
\end{proof}

\begin{fact}[\cite{YamadaOguraAPSMNFAO}]\label{fact:special.Fejer}
Let a sequence $(\bm{x}_n)_{n\in\Natural}\subset\Real^L$, and a
closed convex set $C\subset\Real^L$. Assume that
\begin{equation*}
\exists\kappa>0:\ \forall \bm{f}\in C,\ \forall n\in\Natural, \quad
\kappa\norm{\bm{x}_{n+1} - \bm{x}_n}^2 \leq \norm{\bm{x}_n - \bm{f}}^2 -
\norm{\bm{x}_{n+1} - \bm{f}}^2.
\end{equation*}
Assume, also, that there exists a hyperplane $\Pi$ such that the
relative interior of the set $C$ with respect to $\Pi$ is nonempty,
i.e., $\relinterior_{\Pi}C\neq \emptyset$. Then, $\exists
\bm{x}_*\in\Real^L:\ \bm{x}_*=\lim_{n\rightarrow\infty}
\bm{x}_n$.

Here, given any $\Upsilon\subset\Real^L$, $\relinterior_{\Upsilon}C :=
\{\bm{y}\in\Real^L:\ \exists\rho>0, B(\bm{y},\rho) \cap \Upsilon
\subset C\}$. As a byproduct of this definition, the interior of $C$
is defined as $\interior C:= \relinterior_{\Real^L} C$. Hence, it
becomes clear that if $\interior C\neq \emptyset$, then we can always
find a hyperplane $\Pi\subset\Real^L$ such that $\relinterior_{\Pi}
C\neq \emptyset$. This fact will be used in the proof of
Theorem~\ref{thm:the.analysis}.\ref{thm:liminf}.\qed
\end{fact}

\begin{fact}[\cite{YamadaOguraAPSMNFAO}]\label{fact:exterior.point}
Let $C\subset\Real^L$ be a nonempty closed convex set. Assume also an
 $\mathring{\bm{f}}\in\interior C$, i.e., $\exists \rho>0$ such that
 $B(\mathring{\bm{f}},\rho)\subset C$. Assume, now, an
 $\bm{x}\in\Real^L\setminus C$, and a $t\in(0,1)$ such that
 $\mathring{\bm{f}} + t(\bm{x}-\mathring{\bm{f}})\notin C$. Then,
$d(\bm{x},C) > \rho \frac{1-t}{t}$.\qed
\end{fact}

\begin{lemma}\label{lem:bounded.subgrads}
The set of all subgradients of the collection of convex functions
 $(\Theta_n)_{n\in\Natural}$, defined in \eqref{special.Theta_n}, is
 bounded, i.e., $\forall n\in\Natural$, $\forall \bm{x}\in\Real^L$,
 $\norm{\Theta_n'(\bm{x})}\leq 1$.\qed
\end{lemma}

\begin{proof}
Fix arbitrarily an $n\in\Natural$. Here we deal only with the case
$\mathcal{I}_n\neq\emptyset$, since otherwise, the function $\Theta_n$
becomes everywhere zero, and for such a function,
 Lemma~\ref{lem:bounded.subgrads} holds trivially.

By \eqref{special.Theta_n}, Fact~\ref{fact:partial.distance}, and some
 calculus on subdifferentials \cite{Hiriart}, we obtain that $\forall
 \bm{x}\in\Real^L$, the norm of any subgradient $\Theta_n'(\bm{x})$
 satisfies the following:
\begin{align*}
\norm{\Theta_n'(\bm{x})} & = \norm{\sum_{j\in\mathcal{I}_n}
\frac{\omega_j^{(n)} d(\bm{h}_n,S_j[\epsilon])}{L_n}
d'(\bm{x},S_j[\epsilon])} \\
& = \norm{\sum_{j\in\mathcal{J}_n}
\frac{\omega_j^{(n)} d(\bm{h}_n,S_j[\epsilon])}{L_n} d'(\bm{x},S_j[\epsilon])}\\
& \leq \sum_{j\in\mathcal{J}_n:\ \bm{x}\notin
S_j[\epsilon]} \frac{\omega_j^{(n)}
d(\bm{h}_n,S_j[\epsilon])}{L_n}\norm{d'(\bm{x},S_j[\epsilon])}\\
& \hspace{50pt} + \sum_{j\in\mathcal{J}_n:\ \bm{x}\in
 S_j[\epsilon]} \frac{\omega_j^{(n)}
d(\bm{h}_n,S_j[\epsilon])}{L_n}\norm{d'(\bm{x},S_j[\epsilon])}\\
& \leq \sum_{j\in\mathcal{J}_n:\ \bm{x}\notin
 S_j[\epsilon]} \frac{\omega_j^{(n)} d(\bm{h}_n,S_j[\epsilon])}{L_n}
 \frac{\norm{\bm{x}- P_{S_j[\epsilon]}(\bm{x})}}{d(\bm{x},S_j[\epsilon])} +
 \sum_{j\in\mathcal{J}_n:\ \bm{x}\in
 S_j[\epsilon]} \frac{\omega_j^{(n)} d(\bm{h}_n,S_j[\epsilon])}{L_n} \\
& = \sum_{j\in\mathcal{J}_n:\ \bm{x}\notin
S_j[\epsilon]} \frac{\omega_j^{(n)} d(\bm{h}_n,S_j[\epsilon])}{L_n} +
\sum_{j\in\mathcal{J}_n:\ \bm{x}\in S_j[\epsilon]} \frac{\omega_j^{(n)}
d(\bm{h}_n,S_j[\epsilon])}{L_n} = 1.
\end{align*}
This establishes Lemma~\ref{lem:bounded.subgrads}.
\end{proof}

\subsection{The proof of Theorem~\ref{thm:the.analysis}.}

\begin{enumerate}

\item Assumption~\ref{ass:nonempty.intersection},
  Definition~\ref{def:subgrad.projection}, and
  \eqref{condition.4.lev.Theta.n} suggest that \eqref{basic.recursion}
  can be equivalently written as follows: $\forall n\geq z_0$,
  $\bm{h}_{n+1} = P_{B_{\ell_1}[\bm{w}_n,\delta]}
  T_{\Theta_n}^{(\lambda_n)}(\bm{h}_n)$, where
  $T_{\Theta_n}^{(\lambda_n)}$ stands for the relaxed subgradient
  projection mapping with respect to $\Theta_n$. Notice here that
  $\forall n\geq z_0$, $\lev\Theta_n=\bigcap_{j\in\mathcal{I}_n}
  S_j[\epsilon] \supset \bigcap_{j\in\mathcal{J}_n}
  S_j[\epsilon]$. Thus, by Assumption~\ref{ass:nonempty.intersection}
  and Lemma~\ref{lem:compose.P.and.T}, we have that $\forall n\geq
  z_0, \forall \bm{f}\in\Omega$,
\begin{align}
0 & \leq \frac{2-\lambda_n}{2} \norm{\bm{h}_n -
  \bm{h}_{n+1}}^2 = \frac{2-\lambda_n}{2}
\norm{\bm{h}_n - P_{B_{\ell_1}[\bm{w}_n,\delta]}
 T_{\Theta_n}^{(\lambda_n)}(\bm{h}_n)}^2 \nonumber\\
& \leq \norm{\bm{h}_n - \bm{f}}^2 - \norm{P_{B_{\ell_1}[\bm{w}_n,\delta]}
 T_{\Theta_n}^{(\lambda_n)}(\bm{h}_n) - \bm{f}}^2 = \norm{\bm{h}_n - \bm{f}}^2  -
\norm{\bm{h}_{n+1} - \bm{f}}^2\label{strongly.attracting.Omega_n}\\
& \Rightarrow \norm{\bm{h}_{n+1} - \bm{f}} \leq \norm{\bm{h}_n -
  \bm{f}}.\label{monotonicity}
\end{align}
If we apply $\inf_{\bm{f}\in\Omega}$ on both sides of
\eqref{monotonicity},  we establish our original claim.

\item The next claim is to show that under
      Assumption~\ref{ass:nonempty.intersection}, the sequence
  $(\norm{\bm{h}_n-\bm{f}})_{n\in\Natural}$ converges $\forall
      \bm{f}\in\Omega$. To this end, fix arbitrarily
      $\bm{f}\in\Omega$. By \eqref{monotonicity}, the
  sequence $(\norm{\bm{h}_n-\bm{f}})_{n\geq z_0}$ is non-increasing,
  and bounded below. Hence, it is convergent. This establishes the claim.

Next we will show that under Assumption~\ref{ass:nonempty.intersection},
      the set of all cluster points of the sequence $(\bm{h}_n)_{n\in\Natural}$
  is nonempty, i.e., $\mathfrak{C}((\bm{h}_n)_{n\in\Natural})\neq
  \emptyset$.

  We will first show that the sequence $(\bm{h}_n)_{n\in\Natural}$ is
      bounded. This can be easily verified as follows; fix arbitrarily
      an $\bm{f}\in\Omega$ and notice that $\forall n\geq z_0$,
      $\norm{\bm{h}_n} \leq \norm{\bm{h}_n-\bm{f}} + \norm{\bm{f}} \leq
      \norm{\bm{h}_{z_0} -\bm{f}} + \norm{\bm{f}}$. Define now $D:=
      \max\{\norm{\bm{h}_{z_0} - \bm{f}} +
      \norm{\bm{f}},\norm{\bm{h}_0},\ldots, \norm{\bm{h}_{z_0-1}}\}$,
      which clearly implies that $\forall n\in\Natural$,
      $\norm{\bm{h}_n}\leq D$. Since $(\bm{h}_n)_{n\in\Natural}$ is
      bounded, there exists a subsequence of $(\bm{h}_n)_{n\in\Natural}$
      which converges to an $\tilde{\bm{h}}_*\in\Real^L$
      (Bolzano-Weierstrass Theorem). Hence, $\tilde{\bm{h}}_*\in
      \mathfrak{C}((\bm{h}_n)_{n\in\Natural})\neq \emptyset$. This
      establishes the claim.

Let Assumptions~\ref{ass:nonempty.intersection} and
      \ref{ass:lambda.epsilon} hold true. Then, we will show that
      $\lim_{n\rightarrow\infty} \Theta_n(\bm{h}_n)=0$. First, we will
      prove that
\begin{equation}
\lim_{n\rightarrow\infty}
 \frac{\Theta_{n}(\bm{h}_n)}{\norm{\Theta'_n(\bm{h}_n)}}=0.
 \label{fraction.goes.to.zero}
\end{equation}
We will show this by deriving a contradiction. To this end, assume that
there exists a $\delta>0$ and a subsequence $(n_k)_{k\in\Natural}$ such
that $\forall k\in\Natural$,
$\frac{\Theta_{n_k}(\bm{h}_{n_k})}{\norm{\Theta'_{n_k}(\bm{h}_{n_k})}}
\geq \delta$. We can always choose a sufficiently large $k_0$ such that
$\forall k\geq k_0$, $n_{k}\geq z_0$.

Let, now, any $\bm{f}\in\Omega$, and recall that $\Omega\subset
B_{\ell_1}[\bm{w}_{n_k},\delta]$, $\forall k\geq k_0$. Then, verify that
the following holds true $\forall k\geq k_0$:
\begin{align}
& \norm{\bm{h}_{n_k+1} - \bm{f}}^2  = \norm{P_{B_{\ell_1}[\bm{w}_{n_k},\delta]}
\left(\bm{h}_{n_k} - \lambda_{n_k}
\frac{\Theta_{n_k}(\bm{h}_{n_k})}{\norm{\Theta_{n_k}'(\bm{h}_{n_k})}^2}
\Theta_{n_k}'(\bm{h}_{n_k}) \right) - \bm{f}}^2\nonumber\\
&\leq \norm{\bm{h}_{n_k} - \lambda_{n_k}
\frac{\Theta_{n_k}(\bm{h}_{n_k})}{\norm{\Theta_{n_k}'(\bm{h}_{n_k})}^2}
\Theta_{n_k}'(\bm{h}_{n_k}) - \bm{f}}^2\nonumber\\
& = \norm{\bm{h}_{n_k} - \bm{f}}^2 + \lambda_{n_k}^2
\frac{\Theta_{n_k}^2(\bm{h}_{n_k})}{\norm{\Theta_{n_k}'(\bm{h}_{n_k})}^2}
- 2\lambda_{n_k}
\frac{\Theta_{n_k}(\bm{h}_{n_k})}{\norm{\Theta_{n_k}'(\bm{h}_{n_k})}^2}
{\Theta_{n_k}'(\bm{h}_{n_k})}^T(\bm{h}_{n_k} - \bm{f}),
\label{before.using.subgrad}
\end{align}
where \eqref{P.strongly.attracting} was used for
$P_{B_{\ell_1}[\bm{w}_{n_k},\delta]}$ in order to derive the previous
      inequality. By the definition of the subgradient, we have that
${\Theta_{n_k}'(\bm{h}_{n_k})}^T(\bm{f}-\bm{h}_{n_k}) +
\Theta_{n_k}(\bm{h}_{n_k}) \leq \Theta_{n_k}(\bm{f}) = 0$. If we merge
this into \eqref{before.using.subgrad}, we obtain the following:
\begin{align*}
\norm{\bm{h}_{n_k+1} - \bm{f}}^2 & \leq \norm{\bm{h}_{n_k} - \bm{f}}^2 +
 \lambda_{n_k}^2
 \frac{\Theta_{n_k}^2(\bm{h}_{n_k})}{\norm{\Theta_{n_k}'(\bm{h}_{n_k})}^2}
 - 2 \lambda_{n_k}
 \frac{\Theta_{n_k}^2(\bm{h}_{n_k})}{\norm{\Theta_{n_k}'(\bm{h}_{n_k})}^2}
 \\
& = \norm{\bm{h}_{n_k} - \bm{f}}^2 -
 \lambda_{n_k} (2-\lambda_{n_k})
 \frac{\Theta_{n_k}^2(\bm{h}_{n_k})}{\norm{\Theta_{n_k}'(\bm{h}_{n_k})}^2}.
\end{align*}
This, in turn, implies that
\begin{equation}
\forall k\geq k_0,\quad 0< (\epsilon'' \delta)^2 \leq \lambda_{n_k}
(2-\lambda_{n_k})
 \frac{\Theta_{n_k}^2(\bm{h}_{n_k})}{\norm{\Theta_{n_k}'(\bm{h}_{n_k})}^2}
  \leq \norm{\bm{h}_{n_k} - \bm{f}}^2 - \norm{\bm{h}_{n_k+1} -
 \bm{f}}^2.\label{root.of.contradiction}
\end{equation}
However, as we have already shown before, $(\norm{\bm{h}_n - \bm{f}})_{n\in\Natural}$ is
convergent, and hence it is a Cauchy sequence. This implies that
      $\lim_{k\rightarrow\infty} (\norm{\bm{h}_{n_k} - \bm{f}}^2 -
      \norm{\bm{h}_{n_k+1} - \bm{f}}^2)=0$, which apparently contradicts
      \eqref{root.of.contradiction}. In other words,
      \eqref{fraction.goes.to.zero} holds true.

Notice, now, that for all those
  $n\in\Natural$ such that $\Theta_n'(\bm{h}_n)\neq 0$, we have by
      Lemma~\ref{lem:bounded.subgrads} that
\begin{equation}
\Theta_n(\bm{h}_n) = \norm{\Theta_n'(\bm{h}_n)}
\frac{\Theta_n(\bm{h}_n)}{\norm{\Theta_n'(\bm{h}_n)}} \leq
\frac{\Theta_n(\bm{h}_n)}{\norm{\Theta_n'(\bm{h}_n)}}.
\label{bounded.subgradient.prio.taking.the.limit}
\end{equation}
Notice, also, here that for all those $n\in\Natural$ such that
$\Theta_n'(\bm{h}_n)= 0$, it is clear by the well-known fact
      $\bm{0} \in\partial\Theta_n(\bm{h}_n) \Leftrightarrow
      \bm{h}_n\in\argmin\{\Theta_n(\bm{x}):\ \bm{x}\in\Real^L\}$ that
      $\Theta_n(\bm{h}_n)=0$. Take $\lim_{n\rightarrow\infty}$ on both
      sides of \eqref{bounded.subgradient.prio.taking.the.limit}, and
      use \eqref{fraction.goes.to.zero} to establish our original claim.

Let now Assumption~\ref{ass:nonempty.intersection} holds true. Then we
      show that there exists a $D>0$ such that $\forall n\in\Natural$,
      $L_n\leq D$. Notice, that $\forall n\geq z_0$, $\forall
      j\in\mathcal{J}_n$, $\forall \bm{f}\in\Omega$,
\begin{align*}
d(\bm{h}_n,S_j[\epsilon])  &= \norm{\bm{h}_n -
 P_{S_j[\epsilon]}(\bm{h}_n)}
\leq \norm{\bm{h}_n - \bm{f}} + \norm{\bm{f} -
 P_{S_j[\epsilon]}(\bm{h}_n)}\\
& \leq 2 \norm{\bm{h}_n - \bm{f}} \leq 2 \norm{\bm{h}_{z_0} - \bm{f}},
\end{align*}
where we have used \eqref{P.strongly.attracting} and the monotonicity of
the sequence $(\norm{\bm{h}_n-\bm{f}})_{n\geq z_0}$. Then, by the
      definition of $L_n$,
\begin{equation*}
\forall n\geq z_0, \quad L_n = \sum_{j\in\mathcal{J}_n} \omega_j^{(n)}
d(\bm{h}_n,S_j[\epsilon]) \leq 2 \sum_{j\in\mathcal{J}_n}
\omega_j^{(n)} \norm{\bm{h}_{z_0} - \bm{f}} = 2 \norm{\bm{h}_{z_0} - \bm{f}}.
\end{equation*}
Choose, now, any $D > \max\{2\norm{\bm{h}_{z_0} - \bm{f}}, L_0,
\ldots, L_{z_0-1}\}\geq 0$, and notice that for such a $D$ the claim
holds true.

Let Assumptions~\ref{ass:nonempty.intersection},
      \ref{ass:lambda.epsilon}, and \ref{ass:omega} hold true. By
      \eqref{special.Theta_n}, we observe that
\begin{align*}
\frac{D}{\check{\omega}} \Theta_n(\bm{h}_n) & =
\frac{D}{\check{\omega}} \sum_{j\in\mathcal{J}_n} \frac{\omega_j^{(n)}
  d^2(\bm{h}_n,S_j[\epsilon])}{L_n} \geq \frac{D}{\check{\omega}}
\sum_{j\in\mathcal{J}_n} \frac{\omega_j^{(n)}
  d^2(\bm{h}_n,S_j[\epsilon])}{D}\\ & \geq \frac{D}{\check{\omega}}
\frac{\check{\omega}}{D} \sum_{j\in\mathcal{J}_n}
d^2(\bm{h}_n,S_j[\epsilon]) \geq
\max\{d^2(\bm{h}_n,S_j[\epsilon]):\ j\in
\mathcal{J}_n\}.\label{max.distance}
\end{align*}
Hence, if we take $\lim_{n\rightarrow\infty}$ on both sides of the
previous inequality, we establish
      Theorem~\ref{thm:the.analysis}.\ref{thm:distance.2.hyperslabs}.

\item Here we establish
      Theorem~\ref{thm:the.analysis}.\ref{thm:distance.2.l1.balls}. Let
      Assumptions~\ref{ass:nonempty.intersection} and
      \ref{ass:lambda.epsilon} hold true. We utilize first
      \eqref{P.strongly.attracting} and then
      \eqref{T.strongly.attracting} in order to obtain the following:
      $\forall \bm{f}\in\Omega$,
\begin{align*}
& \norm{(I-P_{B_{\ell_1}[\bm{w}_n,\delta]})
  (T_{\Theta_n}^{(\lambda_n)}(\bm{h}_n))}^2  
 \leq
 \norm{T_{\Theta_n}^{(\lambda_n)}(\bm{h}_n) - \bm{f}}^2 -
 \norm{P_{B_{\ell_1}[\bm{w}_n,\delta]}T_{\Theta_n}^{(\lambda_n)}(\bm{h}_n)
   - \bm{f}}^2 \\
   & = \norm{T_{\Theta_n}^{(\lambda_n)}(\bm{h}_n) -
   \bm{f}}^2 - \norm{\bm{h}_{n+1} - \bm{f}}^2 \\
   & \leq \norm{\bm{h}_n
   - \bm{f}}^2 - \frac{2-\lambda_n}{\lambda_n} \norm{\bm{h}_n -
 T_{\Theta_n}^{(\lambda_n)}(\bm{h}_n)}^2 
 -\norm{\bm{h}_{n+1} - \bm{f}}^2
 \leq \norm{\bm{h}_n - \bm{f}}^2 -
 \norm{\bm{h}_{n+1} - \bm{f}}^2.
\end{align*}
Take $\lim_{n\rightarrow\infty}$ on both sides of this inequality and
      recall that the sequence $(\norm{\bm{h}_n-\bm{f}})_{n\in\Natural}$
      is convergent, and thus Cauchy, in order to obtain
\begin{equation}
\lim_{n\rightarrow\infty} \norm{(I-P_{B_{\ell_1}[\bm{w}_n,\delta]})
(T_{\Theta_n}^{(\lambda_n)}(\bm{h}_n))} = \lim_{n\rightarrow\infty}
d(T_{\Theta_n}^{(\lambda_n)}(\bm{h}_n),B_{\ell_1}[\bm{w}_n,\delta]) =
0. \label{aux.4.dist.l1.ball.goes.zero}
\end{equation}

Moreover, notice that for all $n\geq z_0$ such that
$\bm{h}_n\notin \lev\Theta_n$, by \eqref{condition.4.lev.Theta.n} we
      obtain that
\begin{equation*}
\norm{\bm{h}_n - T_{\Theta_n}^{(\lambda_n)}(\bm{h}_n)}
= \norm{\bm{h}_n -
 \bm{h}_n + \lambda_n
\frac{\Theta_n(\bm{h}_n)}{\norm{\Theta_n'(\bm{h}_n)}^2}
\Theta_n'(\bm{h}_n)}
= \lambda_n
\frac{\Theta_n(\bm{h}_n)}{\norm{\Theta_n'(\bm{h}_n)}} \leq 2
 \frac{\Theta_n(\bm{h}_n)}{\norm{\Theta_n'(\bm{h}_n)}}.
\end{equation*}
Take $\lim_{n\rightarrow\infty}$ on both sides of this inequality, and
recall \eqref{fraction.goes.to.zero} to easily verify that
\begin{equation}
\lim_{n\rightarrow\infty} \norm{\bm{h}_n -
T_{\Theta_n}^{(\lambda_n)}(\bm{h}_n)} =
0.\label{sequential.demiclosed.subgrad.2}
\end{equation}

Notice, now, that $\forall \bm{f}\in B_{\ell_1}[\bm{w}_n,\delta]$, the
triangle inequality implies that
\begin{equation*}
\norm{\bm{h}_n - \bm{f}} \leq \norm{\bm{h}_n -
  T_{\Theta_n}^{(\lambda_n)}(\bm{h}_n)} +
\norm{T_{\Theta_n}^{(\lambda_n)}(\bm{h}_n) -\bm{f}}.
\end{equation*}
If we take $\inf_{f\in B_{\ell_1}[\bm{w}_n,\delta]}$ on both sides of the previous
inequality, then
\begin{equation*}
\forall n\in\Natural, \quad d(\bm{h}_n,B_{\ell_1}[\bm{w}_n,\delta])
\leq \norm{\bm{h}_n -  T_{\Theta_n}^{(\lambda_n)}(\bm{h}_n)} +
d(T_{\Theta_n}^{(\lambda_n)}(\bm{h}_n),B_{\ell_1}[\bm{w}_n,\delta]).
\end{equation*}
Take, now, $\lim_{n\rightarrow\infty}$ on both sides of this
inequality, and use \eqref{aux.4.dist.l1.ball.goes.zero} and
\eqref{sequential.demiclosed.subgrad.2} to establish
Theorem~\ref{thm:the.analysis}.\ref{thm:distance.2.l1.balls}.

\item Next, let Assumptions~\ref{ass:nonempty.intersection},
  \ref{ass:lambda.epsilon}, and \ref{ass:int.Omega} hold
  true. By \eqref{strongly.attracting.Omega_n} notice that $\forall
      n\geq z_0$, $\forall \bm{f}\in\Omega$,
\begin{equation*}
\frac{\epsilon''}{2} \norm{\bm{h}_n -
  \bm{h}_{n+1}}^2 \leq \frac{2-\lambda_n}{2} \norm{\bm{h}_n -
  \bm{h}_{n+1}}^2  \leq \norm{\bm{h}_n - \bm{f}}^2  - \norm{\bm{h}_{n+1}
  - \bm{f}}^2.
\end{equation*}
This and Fact~\ref{fact:special.Fejer} suggest that $\exists
      \tilde{\bm{h}}_*\in\Real^L:\
      \lim_{n\rightarrow\infty}\bm{h}_n=\tilde{\bm{h}}_*$, i.e.,
      $\{\tilde{\bm{h}}_*\}= \mathfrak{C}((\bm{h}_n)_{n\in\Natural})$.

Now, in order to establish
Theorem~\ref{thm:the.analysis}.\ref{thm:liminf}, let
Assumptions~\ref{ass:nonempty.intersection}, \ref{ass:lambda.epsilon},
\ref{ass:int.Omega}, and \ref{ass:omega} hold true. Notice that the
existence of the unique cluster point $\tilde{\bm{h}}_*$ is guaranteed
by the previously proved claim. To prove
Theorem~\ref{thm:the.analysis}.\ref{thm:liminf}, we will use
contradiction. In other words, assume that
$\tilde{\bm{h}}_*\notin\overline{\liminf_{n\rightarrow\infty}}
\bigcap_{j\in\mathcal{J}_n} S_j[\epsilon]$. This clearly implies that
$\tilde{\bm{h}}_*\notin\liminf_{n\rightarrow\infty}
\bigcap_{j\in\mathcal{J}_n} S_j[\epsilon]$. For the sake of compact
notations, we define here $\forall n\in\Natural$, $\Psi_n :=
\bigcap_{j\in\mathcal{J}_n} S_j[\epsilon]$.

Note that the set $\overline{\liminf_{n\rightarrow\infty}}
      \Psi_n$ is convex. This comes from the fact that
      $\Psi_n$ and $\bigcap_{m\geq n} \Psi_m$ are convex,
      $\forall n\in\Natural$, and that $\forall n\in\Natural$,
      $\bigcap_{m\geq n} \Psi_m \subset \bigcap_{m\geq n+1} \Psi_m$.

Since by our initial assumption $\interior \bigcap_{n\geq z_0} \Psi_n
\neq \emptyset$, we can always find an $\mathring{\bm{f}}$ and a
$\rho>0$ such that $B(\mathring{\bm{f}},\rho)\subset \bigcap_{n\geq
  z_0} \Psi_n$. Hence,
\begin{equation}
\forall n\geq z_0, \quad B(\mathring{\bm{f}},\rho) \subset
\Psi_n.\label{ball.in.lev.Theta_n}
\end{equation}

Notice, here, that $\mathring{\bm{f}}\in \bigcap_{n\geq z_0}
\Psi_n \subset \bigcup_{n\in\Natural} \bigcap_{m\geq n}
\Psi_m =: \liminf_{n\rightarrow\infty} \Psi_n \subset
\overline{\liminf_{n\rightarrow\infty}} \Psi_n$. Using this, our initial
      assumption on $\tilde{\bm{h}}_*$, and the fact that
$\overline{\liminf_{n\rightarrow\infty}} \Psi_n$ is closed and
convex, then we can always find a $t\in(0,1)$ such that $\bm{f}_t:=
\mathring{\bm{f}}+ t(\tilde{\bm{h}}_*- \mathring{\bm{f}})\notin
\overline{\liminf_{n\rightarrow\infty}} \Psi_n$. This implies, by
the definition of $\liminf_{n\rightarrow\infty} \Psi_n$, that
\begin{equation}
\forall n\geq z_0,\quad \bm{f}_t \notin \bigcap_{m\geq
    n}\Psi_m. \label{u_t.notin.intersection}
\end{equation}

Now, since $\lim_{n\rightarrow\infty} \bm{h}_n=\tilde{\bm{h}}_*$,
there exists a $z_1\in\Natural$ such that $\forall n\geq z_1$,
$\norm{\tilde{\bm{h}}_* - \bm{h}_n} < \frac{\rho(1-t)}{2t}$. If we set
$n$ equal to $\max\{z_0,z_1\}$ in \eqref{ball.in.lev.Theta_n} and
\eqref{u_t.notin.intersection}, then we readily verify that $\exists
n_0\in\Natural$ such that $n_0\geq \max\{z_0,z_1\}$,
$B(\mathring{\bm{f}},\rho)\subset \Psi_{n_0}=
\bigcap_{j\in\mathcal{J}_{n_0}} S_j[\epsilon]$ and $\bm{f}_t\notin
\Psi_{n_0}$. The result $\bm{f}_t\notin \Psi_{n_0}$ is obviously
equivalent to: $\exists j_0\in \mathcal{J}_{n_0}$ such that
$\bm{f}_t\notin S_{j_0}[\epsilon]$. Also, notice that
$B(\mathring{\bm{f}},\rho)\subset S_{j_0}[\epsilon]$. Hence,
Fact~\ref{fact:exterior.point} suggests that
$d(\tilde{\bm{h}}_*,S_{j_0}[\epsilon]) > \frac{\rho(1-t)}{t}$.

Using the triangle inequality $\norm{\tilde{\bm{h}}_*- \bm{f}}\leq
\norm{\tilde{\bm{h}}_*- \bm{h}_{n_0}} + \norm{\bm{h}_{n_0} - \bm{f}}$,
$\forall \bm{f}\in S_{j_0}[\epsilon]$, we obtain the following:
$d(\bm{h}_{n_0},S_{j_0}[\epsilon]) \geq
d(\tilde{\bm{h}}_*,S_{j_0}[\epsilon]) - \norm{\tilde{\bm{h}}_*-
  \bm{h}_{n_0}} > \frac{\rho(1-t)}{t} - \frac{\rho(1-t)}{2t} =
\frac{\rho(1-t)}{2t}=: \gamma>0$. This clearly implies that
$\max\{d(\bm{h}_{n_0},S_{j}[\epsilon]):\ j\in\mathcal{J}_{n_0}\} \geq
\gamma>0$. Set, now, $n$ equal to $n_0+1$ in
\eqref{ball.in.lev.Theta_n} and \eqref{u_t.notin.intersection}, and
verify, as we did before, that $\exists n_1\in\Natural$ such that
$\max\{d(\bm{h}_{n_1},S_{j}[\epsilon]):\ j\in\mathcal{J}_{n_1}\} \geq
\gamma>0$. Going on this way, we can construct a sequence
$(\bm{h}_{n_k})_{k\in\Natural}$ such that $\forall k\in\Natural$,
$\max\{d(\bm{h}_{n_k},S_{j}[\epsilon]):\ j\in\mathcal{J}_{n_k}\} \geq
\gamma>0$. However, this contradicts
Theorem~\ref{thm:the.analysis}.\ref{thm:distance.2.hyperslabs}.  Since
we have reached a contradiction, this means that our initial
assumption is wrong, and that
$\tilde{\bm{h}}_*\in\overline{\liminf_{n\rightarrow\infty}}
\bigcap_{j\in\mathcal{J}_n}S_j[\epsilon]$.

If we follow exactly the same procedure, as we did before, for the
case of the sequence of sets
$(B_{\ell_1}[\bm{w}_n,\delta])_{n\in\Natural}$, then we obtain also
$\tilde{\bm{h}}_*\in\overline{\liminf_{n\rightarrow\infty}}
B_{\ell_1}[\bm{w}_n,\delta]$.

\end{enumerate}

\bibliographystyle{ieeebib}

\end{document}